\documentstyle[12pt,epsf,aps]{revtex}
\begin{document} 
\draft
\title{Glueball-quarkonia content and decay of scalar-isoscalar mesons\\[2ex]}
\author{M. Strohmeier-Pre\v{s}i\v{c}ek$^{1}$, T. Gutsche$^{1}$, R. Vinh 
Mau$^{2}$ and  Amand Faessler$^{1}$\\[4ex]}
\address{
$^1$ Institut f\"ur Theoretische Physik, Universit\"at T\"ubingen,
Auf der Morgenstelle 14, \\
D-72076 T\"ubingen, Germany \\[2ex]
$^2$ Laboratoire de Physique Th\'eorique des Particules \'El\'ementaires,\\
Universit\'e P. et M. Curie, 4 Place Jussieu, 75252 Paris Cedex 05, France}
\maketitle
\begin{center}
{\bf Abstract}
\end{center}
\begin{abstract}
We investigate the hadronic two-body decay modes of the scalar-isoscalar  
$f_0(1370), ~f_0(1500)$ and 
$f_0(1710)$  states as  resulting from the mixture of the lowest lying 
scalar glueball with the isoscalar states of the 
ground state  $^3P_0$ $Q\bar Q$   nonet. In 
the decay analysis we take into account the direct coupling 
of the quarkonia and glueball components of the $f_0$ states to the 
quarkonia components of the two-meson final state with the  decay dynamics 
 inspired by the strong coupling limit of QCD. We calculate partial 
decay widths for the $f_0$ states in the proposed three-state mixing schemes 
and discuss their compatibility with the observed decay features. 
Finally, we determine the glueball-quarkonia content of the $f_0$ states 
from a  detailed fit 
 to experimental decay data of  $f_0(1500)$ and give predictions 
for the partial decay widths of $f_0(1370)$ and $f_0(1710)$, providing thus
a sensitive test of the proposed mixing scheme. 
\end{abstract}

\newpage
\section{Introduction}
The identification of mesons beyond the conventional valence quark-antiquark
$(Q\bar Q)$ pair 
configuration is one of the most challenging tasks in low-energy 
meson spectroscopy. \\
For instance, 
as a consequence of gluon self-couplings in QCD the usual $Q\bar Q$  
spectrum is expected to be supplemented by mesonic states with 
pure gluonic composition, that is  glueballs \cite{clos}.\\
Lattice QCD,  in the quenched approximation,
predicts the lightest glueball state
$(G_0)$ to be a scalar $(J^{PC}=0^{++})$ lying in the mass
 range of 1.4 - 1.8 GeV \cite{bal,tep}.
Since the observed  isovector and isodoublet states $a_0(1450)$ and $K_0(1430)$ 
set the natural 
 mass scale of the ground state scalar meson nonet, 
the isoscalar $Q\bar Q$ partners
 $n \bar n \equiv  \frac{1}{\sqrt{2}}(u\bar{u}+d\bar{d})$
and $s \bar s$ are expected to lie in a mass region close    to the 
 predicted ground state  glueball.
The vicinity of  masses for the pure  $G_0$, $n\bar n$ and $s\bar s$ mesonic 
states suggests that the intrusion of the scalar glueball into the 
scalar $Q\bar Q$  meson spectrum 
 is accompanied by significant mixing of $G_0$ 
with the isoscalar members of the $^3P_0$ nonet \cite{ams,weina}.\\
Several glueball-quarkonia mixing schemes \cite{ams,weina,anis,naris}
 have been proposed to account 
for this scenario and to reveal the $G_0/Q\bar Q$ nature of the
 scalar-isoscalar 
states $f_0(1370)$, $f_0(1500)$,  established  
by Crystal Barrel at LEAR \cite{amsler} in proton-antiproton annihilation
 reactions,
 and the
 $f_0(1710)$ \cite{bugg1710}, which is the scalar component of the  
$f_{j=0,2}(1710)$ \cite{pdg}.
In a simplified picture considered in Refs.
 \cite{ams,weina} the scalar glueball ground state $G_0$
mixes with the lowest lying $^3P_0$ isoscalar quarkonium states.
Further admixture of the first radially excited $^3P_0$ nonet is introduced 
in Refs. \cite{anis,naris}. However, the restriction to the glueball 
and near lying $^3P_0$ quarkonia ground states is a good approximation 
 provided that  the excited glueball 
and quarkonia states are high enough in mass.\\ 
Quantitative predictions in the three-state mixing 
schemes of Refs. \cite{ams,weina}
for the glueball-quarkonia content of  
   $f_0(1370)$, $f_0(1500)$ and $f_0(1710)$ 
differ substantially due to orthogonal theoretical 
assumptions concerning the level ordering    
 of the bare states before mixing. While Ref. \cite{ams} uses  
for the input  masses  of the bare states 
$M(n\bar n)< M(G_0)<M(s\bar s)$,   in 
Ref. \cite{weina} the pure glueball 
 lies above the scalar $Q\bar Q$ state. The difference of level 
ordering in the input bare masses
leads to a substantially 
 different $G_0/Q\bar Q$ content of the physical $f_0$ states, 
especially for $f_0(1500)$ and $f_0(1710)$.
While in   Ref. \cite{ams}, the glueball is distributed with nearly equal 
strength over the three $f_0$ states and  $f_0(1710)$ has a large $s \bar s$ 
component,  the analysis
  of Ref. \cite{weina} 
 leads  to  a dominant $G_0$ structure  in  the  $f_0(1710)$, while 
$f_0(1500)$ is a nearly pure $s\bar s$ state.
 The results for the $f_0(1370)$ are comparable in both approaches, 
with   a large $n \bar n$ component  residing in this state. 
The observed
  total width of the  $f_0(1370)$  \cite{amsler}
  and its strong coupling
to the $\pi\pi$ decay channel is in good agreement 
with naive quark model expectations \cite{kok,god} for a $n\bar n$ $^3P_0$
state and seems to confirm its dominant $n\bar n$ structure. 
Therefore, the main difference between the two mixing schemes is the amount 
of  glueball  and $s\bar s$ strength residing in the $f_0(1500)$ 
and $f_0(1710)$  states, respectively.  A further test of the proposed 
mixing schemes, where the glueball ground state
 intrudes in the scalar quarkonium
sector, is the analysis of the decay modes of the observed $f_0$ states.
The decay of $f_0(1500)$ into  two pseudoscalar mesons was studied in  
leading  order of strong coupling QCD \cite{ams}.  
Neglecting possible 
 gluonic components of the isoscalar mesons in the final state, 
 the $ \pi\pi, K\bar K,  \eta\eta$ and $ \eta\eta^{\prime} $decay modes  
  are entirely driven  by the quarkonia components of the 
$f_0(1500)$ state. In this approximation, 
results for  the partial decay widths deduced from $SU(3)$ flavor 
 symmetric  couplings are quite similar for both mixing schemes and 
cannot be discriminated \cite{ams}, e.g. 
the observed  weak $K\bar K$ decay mode is obtained in 
 both schemes by the destructive interference between the  $n\bar n$ 
and $s\bar s$ components. \\
In order to reveal the difference of the proposed three state mixing 
schemes, one should take into account the direct coupling
of the $G_0$ component to the final state mesons.
This mechanism  occurs in the leading of strong coupling QCD \cite{ams} 
by the transition $G_0 \rightarrow G_0 G_0$,
which can produce final states with isoscalar  mesons $(\eta\eta,
\eta\eta^\prime$ and $\sigma\sigma$, where  $\sigma$ is the broad low 
lying $\pi\pi$ S-wave resonance)  with possibly a sizable gluonic 
component.
Independent of a possible small  glueball component in the final state 
isoscalars  is the decay 
 process $G_0 \rightarrow (Q\bar Q) (Q\bar Q)$. This 
 next-to-leading order transition can  contribute to all two-body decay modes 
of the mixed $f_0$ states. 
An   investigation \cite{mitja} of the two-body decay modes 
$\sigma\sigma, \rho\rho$ and $\pi^\ast(1300)\pi$ of the $f_0(1500)$, 
all leading to the $4\pi$ decay channel, indicates the possible presence of a 
sizable direct coupling of the  gluonic component of the $f_0(1500)$ to 
the $4\pi$ decay channels. 
Neglecting a possible $G_0$ component in $\sigma$, 
the leading order decay mechanism predicts \cite{mitja}
the hierarchy of branching ratios BR  with
$BR(\rho\rho)^{>}_{\sim} BR(\pi\pi) ^{>}_{\sim} BR(\sigma\sigma)
 > BR(\pi^*(1300)\pi)$, independent of any particular three-state 
mixing scheme.
The results are  in strong conflict with current  Crystal Barrel data,
where  the  $\rho\rho$ decay mode of the $f_0(1500)$ is weak or even 
absent \cite{thoma} and a  stronger coupling  to the $4\pi$ modes  than to 
the $2\pi$ decay channel \cite{amsler} is required. \\ 
In the present  work, going beyond the lowest order decay mechanism, we
couple 
the $Q\bar Q$ components of the mixed $f_0$ states to the quarkonia 
components of  the final state 
 mesons. In the decay analysis we also take  into account the transition 
$G_0 \rightarrow (Q\bar Q)~(Q\bar Q)$ occurring as a next-to-leading 
order decay mechanism deduced from  strong coupling QCD \cite{ams}. Inclusion 
of the direct coupling of the $G_0$ component of the mixed $f_0$ states 
to the quarkonia content of the final state  mesons   
will lead to contradictory physical consequences in the proposed 
three-state mixing schemes of Refs. \cite{ams,weina}, in particular 
 for the physical  
state  $f_0(1710)$.
Because of the shortcomings of these proposed mixing 
 models, we undertake to extract a three-state mixing scheme based 
on a fit to   the detailed  experimental 
 data of   $f_0(1500)$ two-body decays  
 and give the predictions for partial decay widths 
of the partner states $f_0(1370)$ and $f_0(1710)$.\\
This paper is organised as follows. In Sec. \ref{mixing} we introduce the 
three-state mixing scheme of the lowest lying  scalar glueball 
with the  quarkonia
 states  
and the decay dynamics as suggested by the strong coupling limit of QCD.
In Sec. \ref{quark} we develop the formalism for 
 the decay of the $f_0$ states
 by  its $Q \bar Q$ components in the framework of the pair creation 
or $^3P_0$ model.
 In Sec. \ref{gluon} we construct a digluonium wave function based on cavity QCD  for the effective description of the scalar glueball state. 
Furthermore, we indicate the transition mechanism for the decay 
of the $G_0$ component into $(Q\bar Q)~(Q\bar Q)$.  
The results are presented in Sec. \ref{clo-wein}; 
we first show how distinct decay patterns 
result from the unmixed $Q\bar Q$ and $G_0$ components of 
the $f_0$ states.   
We then give partial  widths for the two-body  decay modes  of the 
$f_0$ states in  
the mixing schemes of Refs. \cite{ams,weina} and confront the results with 
experiment. Finally,   we present a three-state mixing scheme deduced 
from a  detailed fit to the   experimental data  for $f_0(1500)$ decays 
and give predictions for the 
 partial decay  
widths of  $f_0(1370)$ and $f_0(1710)$,  
 providing thus a sensitive test of the proposed mixing scheme. 
A summary and conclusions are given in Sec. \ref{summary}.
\newpage

\section{Three-state mixing and decay of  scalar-isoscalar states  in strong coupling QCD}
\label{mixing}
In the lattice formulation of the 
 strong coupling limit $(g \rightarrow \infty)$  of QCD  \cite{ams,kok},
  glueballs and 
$Q\bar Q$ mesons are noninteracting eigenstates of the QCD Hamiltonian.
Departure  from this limit causes mixing and decay of the 
strong coupling eigenstates  \cite{ams,kok}.\\
To lowest order, the mixing of the scalar glueball $G_0$ and the $^3P_0$ quarkonia states 
$n\bar n=1/\sqrt(2)~(u\bar u+d\bar d)$ and  $s\bar s$   can be introduced  
  by the interaction Hamiltonian 
$H_I$ \cite{weina}
\begin{equation}
(~H_I~)= \left( \begin{array}{ccc}
m_{G_0} & z & \sqrt{2} z\\
z & m_{s\bar s}& 0\\
\sqrt{2}z & 0 & m_{n\bar n}
\end{array} \right), 
\end{equation}
where  $z=<G_0|H_I|s\bar s>=<G_0|H_I|n\bar n>/\sqrt{2}$
 represents the flavor 
independent  mixing strength between the   
glueball and  quarkonium states. 
The masses  of the bare  states $i=G_0, s\bar s, n\bar n$ before 
mixing are denoted by $m_i$.  
Deviation from the $SU(3)$ flavor symmetry is known to be small 
 \cite{weinb} and quarkonia
 mixing is assumed to be a higher 
order perturbation in  the strong coupling eigenstates $|G_0>,~|s\bar s>$ 
and $|n\bar n>$ \cite{ams}.
  All parameters 
of $H_I$ can be taken to be real and positive.\\
$H_I$ possesses three eigenstates $|f_0(M)>$  with
physical masses M 
 , which 
are are given by the linear combinations: 
\begin{equation} \label{mix}
|f_0(M)> = a_{n\bar n}|n\bar n>+a_{s\bar s}|s\bar s>+a_{G_0}|G_0>,
\end{equation}
with   $a_{n\bar n}^2+a_{s\bar s}^2+a_{G_0}^2=1$.
Choosing the bare mass  $m_{n\bar n}$ to be 
 equal to that of the observed 
  isovector state $a_0(1450)$ \cite{pdg}, 
   the remaining parameters in 
$H_I$  can be 
adjusted to give the  masses of the observed resonances
$f_0(1370)$, $f_0(1500)$ and $f_0(1710)$, which fixes the
 quarkonia-glueball
 composition of the physical states \cite{weina}. 
In Ref. \cite{ams} it is assumed that the level spacings of the 
bare states fulfil the relation
 $(m_{G_0}-m_{n\bar n})/2=m_{s\bar s}-m_{G_0}=z$,   
 which again determines the state vector of the 
physical $f_0(M)$ states. In this case $f_0(1500)$ is degenerate in mass with 
the $G_0$ state and the physical  masses M
 of the mixing partners $f_0(1370)$ and 
$f_0(1710)$ depend on the mixing strength z. \\
However, there are sizable
 uncertainties in these procedures, particularly  since 
 the mass of the $f_0(1370)$
 is  poorly determined,  ranging from 1200 - 1500 MeV  \cite{pdg} and 
the exact level spacing of the bare states in unknown. 
 For these  reasons  we will also present an alternative 
strategy by assuming  
 that $f_0(1500)$ is one of the eigenstates of $H_I$ 
and by  extracting the mixing coefficients 
 $a_i$ $(i= n\bar n, s\bar s, G_0)$  directly 
from the experimental two-body decay data.  This 
will allow us to give   the bare masses $m_i$ entering 
  in $H_I$ of Eq. (1),  in terms of the mixing strength z, 
which in turn is restricted by the   physical 
masses  of the partners of the  $f_0(1500)$. The 
resulting glueball-quarkonia composition of these partner states 
, however, is completely independent of z, solely 
determined by the mixing coefficients of the $f_0(1500)$. \\  
In the minimal constituent picture $G_0$ consists of two gluons 
bound in a color singlet configuration with $J^{PC}=0^{++}$.
For finite $g$ the  mixed 
$|f_0(M)>$ decays in lowest order by its  $G_0$ and 
$Q\bar Q$ components. The relevant diagrams are shown in Figs. 
\ref{glueb} and \ref{quarkonia}. The transition
 $G_0 \rightarrow G_0~ G_0$ of Fig. 1
  couples the $G_0$ component of the physical state   $|f_0(M)>$ 
to the $G_0$ component
 of the final state isoscalar mesons. The constituent gluons of  the initial
state decay into two pairs of  low energy gluons which   
involves two three-gluon vertices in the decay mechanism. 
We will neglect this 
transition in our decay analysis by assuming that 
the considered final state isoscalar 
mesons do not contain a $G_0$ component. 
In principle, the transition of Fig. 1 can 
contribute to the decay modes 
$f_0(M) \rightarrow \eta\eta, \eta\eta^\prime$ and $\sigma\sigma$,  
  depending on the unknown overlap of the 
$\eta, \eta^\prime$ and $\sigma$ with 
the corresponding glueball states. Current evidence 
for the relevance of such a process involving 
the pseudoscalars $\eta$ and $\eta^\prime$ is not compelling,  as for 
example discussed in Ref. \cite{ams}. 
For the  scalar $\sigma$ meson a large coupling to the glueball 
state is not excluded \cite{ams,pdg}. Therefore, we should keep 
in mind that the $G_0 \rightarrow G_0~G_0$ transition 
could induce some changes for the $\eta\eta, \eta\eta^\prime$ and 
$\sigma\sigma$ decay modes of the $f_0$ states.\\
The    leading order decay  mechanism  
  $Q\bar Q \rightarrow (Q\bar Q)~(Q\bar Q)$, 
 illustrated in Fig. \ref{quarkonia}, 
leads to the decay  of $|f_0(M)>$ by  its  quarkonia 
components $Q \bar Q$ and is   
familiar from the
 OZI-allowed  meson decay \cite{kok,mitja,meson}. This process  will be 
described in Sec. \ref{quark} in the framework of the $^3P_0$ pair
 creation model \cite{le}.\\
 Here we also take into account the next to leading order
decay mechanism  $G_0 \rightarrow (Q\bar Q)~(Q\bar Q)$
 as indicated  in Fig. \ref{gluea}.
 In this mechanism the $G_0$ component of
the physical $f_0(M)$ states
 couples to the final state   $Q\bar Q$ mesons  by  
conversion of the constituent gluons into quark-antiquark pairs.  
 Sec.  \ref{gluon} is devoted to the description of this process.\\
Neglecting a  possible $G_0$ component 
in the final state isoscalar mesons,  
  the partial decay width $\Gamma_{f_0 \rightarrow BC}$
 for the decay of  $|f_0(M)>$ into a two-meson  state  B C 
 can be written as 
\begin{eqnarray}
\Gamma_{f_0 \rightarrow BC}& = & 2~\pi~ K~ \frac{E_BE_C}{M_{f_0}}~ 
\sum_{l_{BC}}~
    \int d\Omega_K~ |T_{f_0 \rightarrow BC}^{(l_{BC})}|^2 \\
& = & \sum_{l_{BC}}~|\lambda~M_{Q\bar Q \rightarrow BC}^{(l_{BC})}~
+~\gamma^2~M_{G_0 \rightarrow BC}^{(l_{BC})}|^2 ~, \nonumber 
\end{eqnarray}
where $M_{Q\bar Q \rightarrow BC}^{(l_{BC})}$ and 
$M_{G_0 \rightarrow BC}^{(l_{BC})}$ denote 
the transition amplitudes including phase space of the respective 
decay mechanism. The sum extends over the relative orbital angular 
momentum $l_{BC}$ between the mesons B and C. 
With the decay momentum $|\vec K|=K$,  $E_i=\sqrt{M_i^2+\vec K^2}$
is the  energy  of the final state mesons $i=B, C$ with mass $M_i$.
 The strengths of the quark-gluon 
coupling  and  the pair creation amplitude in 
the $^3P_0$ model are factorised out and denoted 
by $\gamma$ and $\lambda$, respectively. 
The relative phase between the two transition amplitudes 
 is  fixed by choosing
 $M_{Q\bar Q \rightarrow BC}^{(l_{BC})}$ and $M_{G_0 \rightarrow BC}^{(l_{BC})}$ to be  real and 
the ratio $\gamma^2/\lambda$ to be  complex with
\begin{equation}
\frac{\gamma^2}{\lambda}~\equiv~\kappa~e^{i\phi}.
\end{equation}
In this case $\Gamma_{f_0 \rightarrow BC}$ is  given by
\begin{eqnarray}
\Gamma_{f_0 \rightarrow BC} & = & \lambda^2 \left\{  M^2_{Q\bar Q \rightarrow BC} +
\kappa^2~   M^2_{G_0 \rightarrow BC} + 2~\kappa ~cos\phi~
\sum_{l_{BC}}~ M^{(l_{BC})}_{Q\bar Q \rightarrow BC}
  M^{(l_{BC})}_{G_0 \rightarrow BC} \right\}.
\end{eqnarray} 
with the summation over $l_{BC}$ included in the definition
 of $M^2_{Q\bar Q (G_0)
 \rightarrow BC}$.
Finally, in order to take into account properly
the available phase space for the decay 
$f_0 \rightarrow BC$ we average over
 the mass spectrum $f(m)$ of broad mesons:
\begin{eqnarray}
\Gamma_{f_0 \rightarrow BC} & = & \int dm_{f_0} dm_B dm_C ~
 \Gamma_{f_0 \rightarrow BC}(m_{f_0}, m_B, m_C)
f(m_{f_0}) f(m_B) f(m_C) \\
f(m) & \propto & \frac{(\Gamma_i/2)^2}{(m-M_i)^2+(\Gamma_i/2)^2} \nonumber
\end{eqnarray}
with a threshold cutoff as in Ref. \cite{mary}. The individual masses 
$M_i$ and widths $\Gamma_i$ of the resonances are taken from the Particle Data 
Group.\\
The mass distribution of the scalar-isoscalar
 $\sigma$ meson is parameterised with $M_\sigma =760~MeV$ and 
$\Gamma_\sigma=640~MeV$ \cite{nag}.  For the $\pi^*(1300)$ we 
 use the resonance  width of $\Gamma_{\pi^*(1300)}=200~MeV$ 
\cite{ulrike}.
 The relationship between $f_0(1370)$ and the broad 
structure around  $1100~MeV$ \cite{amsler} called $f_0(400-1200)$ \cite{pdg} 
is not entirely clear. If the two states are manifestations of a single object,
then   $\Gamma_{f_0(1370)}\approx 700~MeV$ \cite{amsler}.
For two independent states one has
 $\Gamma_{f_0(1370)}=351\pm41~MeV, M_{f_0(1370)}=1360\pm23~MeV$ \cite{amsler} 
and $\Gamma_{f_0(1370)}=230\pm15~MeV, M_{f_0(1370)}=1300\pm15~MeV$ \cite{bugga}.
\newpage
\section{Decay of the quarkonia components}
\label{quark}
The decay of the quarkonia components of the $|f_0(M)>$ states 
into the final state mesons B and  C 
  is  calculated 
in the 
non-relativistic $^3P_0$ or pair creation model \cite{le}. This model 
has a solid foundation in strong coupling QCD \cite{kok,pat,dosch}.
It describes the OZI-allowed decay $Q\bar Q \rightarrow BC$ by  
the creation of a $Q\bar Q$ pair with  quantum numbers 
$(I^G(J^{PC})=0^+(0^{++}))$ out of the hadronic vacuum,  
as indicated in Fig. \ref{quarkonia}.
In the evaluation one  also has to consider  a second 
diagram where the   quark 3 (4) goes into the 
meson C (B). The constituent quarks 
of the initial state are spectators in the transition. The strength 
of this transition is governed by the  dimensionless constant 
$\lambda$, which is related to the pair creation probability.
From detailed fits to tensor meson decay \cite{ams,thomas}
it is concluded that the pair creation mechanism  is flavor independent.\\
The nonperturbative $Q\bar Q$ $^3P_0$ vertex can be defined as 
\begin{equation}
V_{^3P_0}^{(43)}= \lambda~ \delta ( {\vec{p}} _{4}+
{\vec{p}} _{3}) \left[ {\cal Y}_{1\mu}^\ast( {\vec{p}} _{4}
- {\vec{p}} _{3}) \otimes \sigma_{-\mu}^{(43)\dagger} \right]_{00}
{\mathbf{1}} _F^{(43)} {\mathbf{1}} _C^{(43)}~,
\end{equation}
where  $\vec{p}_{4(3)}$ 
are  the momenta   of the  quark (antiquark)  4 (3) and 
 ${\cal Y}_1(\vec p)=|\vec p|~Y_1(\hat p)$.   
 The identity operators $\mathbf 1_F$ and $\mathbf 1_C$ 
project onto singlet states in flavor (F) and color (C)  space of the 
created $Q\bar Q$ pair (43).\\
Using the   harmonic oscillator  ansatz
 $\Psi_{n=0,l=1}(\vec{p_1},\vec{p_2})$, 
 the   wave function of the $Q\bar Q$ component 
 of the mixed $f_0$ can be  
expressed in its centre-of-momentum frame  as:
\begin{eqnarray} \label{a}
\Psi_{f_0}(\vec{p}_1,\vec{p}_2) & = & \delta (
 {\vec{p}}_1 + {\vec{p}}_2 )~
 \left[\chi_{S=1}(12)\otimes
 \Psi_{n=0,l=1} 
({\vec{p}}_1,{\vec{p}}_2)\right]_{J=0}~ \chi_F^{f_0}(12)~\chi_C^{(12)} \nonumber\\
 & = &  \left[ \frac{2R_{Q\bar Q}^5}{3\sqrt{\pi}} \right]^{\frac{1}{2}} \delta (
 {\vec{p}}_1 + {\vec{p}}_2 )
\exp \left\{ - \frac{1}{8} R_{Q\bar Q}^2 ( {\vec{p}}_1-{\vec{p}}_2)^2 \right\}\\
 &  & \cdot \left[ \chi_{S=1}(12) \otimes {\cal Y}_{l=1}({\vec{p}}_1-{\vec{p}}_2) \right]_{J=0} \chi_{F}^{f_0}(12) \chi_C^{(12)} ~, \nonumber
\end{eqnarray}
where $R_{Q\bar Q}$ is the  size parameter for the $Q\bar Q$ component.
The intrinsic spin  $\chi_{S=1}(12)$  of the $Q\bar Q$ pair (12)
is coupled with relative angular momentum $l=1$ to
  the  $J^{PC}=0^{++}$ scalar ground state ($n=0$)  and $\chi_C^{(12)}$ is the color singlet 
wave function. The  $SU(3)$ flavor  part $\chi^{f_0}_F(12)$  is given by
\begin{equation}
\chi^{f_0}_F(12)~=~a_{n\bar n}~|n_1\bar n_2>~+~a_{s\bar s}~|s_1\bar s_2>,
\end{equation}
with the quarkonia  mixing coefficients $a_i$ ($i=n\bar n, s\bar s$) of 
 the three-state mixing scheme introduced in Sec. \ref{mixing}. 
For the final state mesons $i~ (i=B,C)$ with momentum $\vec P_i$, total spin 
$J_i$, relative orbital angular momentum  $l_i$ and intrinsic spin $S_i$
 we use the wave functions 
\begin{equation}
\Psi_{i}(\vec p_l,\vec p_m)~=\delta (
 {\vec{p}}_l + {\vec{p}}_m-\vec P_i )~
[\chi_{S_i}(lm)\otimes\Psi_{n_il_i}
(\vec{p_l},\vec{p_m})]_{J_i}~\chi^i_F(lm)~\chi_C^{(lm)},
\end{equation}
 where  $\vec p_{l(m)}$ are internal
 quark momenta for the quarks $l(m)$.
Explicit forms for the harmonic oscillator wave functions $\Psi_{n_il_i}$ 
used  
 can be found in Appendix A. \\
The transition amplitude $M_{Q\bar Q \rightarrow BC}^{(l_{BC})}$, as
 defined by Eq. (3),
separates into a flavor part $M^F_{Q\bar Q \rightarrow BC}$ 
\begin{equation}
M^F_{Q\bar Q \rightarrow BC}~=~<\chi^B_F(13)\chi^C_F(42)|{\mathbf{1}}_F^{(43)}|\chi_F^{f_0}(12)>
\end{equation}
and the  spin-spatial-color part $M^{(l_{BC})SSC}_{Q\bar Q \rightarrow BC}$
\begin{eqnarray}
M_{Q\bar Q \rightarrow BC}^{(l_{BC})SSC}& = & \sqrt{2\pi K}~
\sqrt{\frac{E_BE_C}{M_{f_0}}}~ \frac{1}{\sqrt{3}}\int\prod_{i=1,...4}d\vec {p_i}
~\delta(\vec{p_4}+\vec{p_3})~ \delta(\vec p_1+\vec p_2) \\
& & \delta(\vec p_1+\vec p_3-\vec K)~  \delta(\vec p_2+\vec p_4+\vec K)~
[\chi_{S_B}^{(13)}\otimes \Psi_{n_Bl_B}(\vec{p_1},\vec{p_3})]^\dagger \nonumber\\
& & [\chi_{S_C}^{(42)}\otimes \Psi_{n_Cl_C}(\vec{p_4},\vec{p_2})]^\dagger
\left[ {\cal Y}_{1\mu}^\ast( {\vec{p}} _{4}
- {\vec{p}} _{3}) \otimes \sigma_{-\mu}^{(43)\dagger} \right]_{00}
[\chi_{S=1}^{(12)}\otimes \Psi_{n=0l=1}(\vec{p_1},\vec{p_2})]_{J=0} \nonumber
\end{eqnarray}
with the decay momentum $\vec K$ and 
  the color matrix element
 $<\chi_C^{(13)}\chi_C^{(42)}|{\mathbf{1}}_C|\chi_C^{(12)}>~=~\sqrt{1/3}$. \\
The amplitude $M_{Q\bar Q \rightarrow BC}^{(l_{BC})SSC}$ 
  is  evaluated analytically. Since there 
exists exhaustive literature \cite{kok,meson,le} on the $^3P_0$ model we just
 cite the 
results  in Appendix B.
Values for the various flavor matrix elements 
are given in  Table \ref{tab1}. The results of Table \ref{tab1}  were 
 obtained by assuming $\chi^{\omega}_F=~|n\bar n>$. Thereby 
we have used the following definitions for the 
  flavor content of the $\eta$ and $\eta^\prime$ mesons: 
\begin{eqnarray}
\psi_F^{\eta}& = & \alpha_{PS}~|n\bar n>~-~\beta_{PS}~|s\bar s> \\
\psi_F^{\eta^\prime}& = & \beta_{PS}~|n\bar s>~+~\alpha_{PS}~|s\bar s>
 \nonumber .
\end{eqnarray}
The quantities $\alpha_{PS}$ and $\beta_{PS}$ are  simply related 
to the pseudoscalar mixing angle $\theta_{P}$ as defined by the Particle Data 
Group \cite{pdg}. For numerical results we use the  
 value of   $\theta_{P}=-~17.3^o$ \cite{ps}  or equivalently
 $\alpha_{PS}~=~0.794$, $\beta_{PS}~=~0.608$. \\
The partial decay widths for the quarkonia decay $Q\bar Q \rightarrow BC$
 are defined  as in  Eq. (5) 
and  depend on the pair creation amplitude 
$\lambda$ and  the quarkonia size parameters $R_{Q\bar Q}$ and  
$R_B=R_C$ for the decay channels considered here.\\ 
 Exhaustive experimental data exist for the hadronic decays of the ground state
tensor meson  nonet and the strategy is to extract the meson size parameters
combined with the $^3P_0$ strength from a full scale decay analysis.
A recent    fit to tensor meson decays \cite{thomas}
yields $\lambda^2=2.11$ and
$R_B=3.65~GeV^{-1}$ for nonstrange S-wave mesons and strange
tensor  mesons, suggesting $R_{a_1}=R_{Q\bar Q}=3.65~GeV^{-1}$.
  S-wave mesons  containing strange quarks  have an effectively 
reduced size  with
$R_B=2.48~GeV^{-1}$ for
$\eta$ and $\eta^\prime$ and  $R_K=2.82~GeV^{-1}$ for K. No flavor symmetry 
breaking at the $^3P_0$ vertex was obtained, 
hence $\lambda$ is independent of the created $Q\bar Q$ flavor. 
Due to   orthogonality of  the $\pi$ and $\pi^*(1300)$
 wave
functions the respective size parameters fulfil 
the relation  $R_{\pi}=R_{\pi^*}$.
The $R_\sigma$ dependence of the quarkonia decay amplitude was shown 
to be small \cite{mitja}. We 
therefore choose $R_\sigma=R_\pi$ for our numerical
 results presented in the forthcoming section.
\newpage
\section{Decay of the glueball component}
\label{gluon}
In this section we consider the two-body 
 decay of the ground state scalar glueball 
($G_0$) in  the next-to-leading 
 order decay mechanism  of  strong coupling QCD \cite{ams}.
 For the decay analysis 
 we  adopt a non-relativistic picture where $G_0$ 
is a bound state of  
 two massive  constituent gluons g. The decay  is proceeding 
via the conversion of the constituent gluons into  $Q\bar Q$  pairs as
illustrated  in Fig. \ref{gluea}. In the evaluation  
a second diagram with different 
quark rearrangement in the final state analogous to quarkonia decay has  also  
 to  be  considered. The
elementary  quark-gluon interaction vertex is   given to lowest order in
the non-relativistic limit by
\begin{equation}
V^{(g_i  Q_l\bar Q_k)}=\gamma~ \delta(\vec q_i-\vec p_l-\vec p_k)~
(\vec{\sigma}_{(lk)}\cdot\vec{\epsilon}_{i})~{\mathbf{1}}_F^{(lk)}~
(\frac{1}{2}\sum_{a=1}^8 \lambda^a_{(lk)}A^a_i)
\end{equation}
with the internal momenta  $\vec q_i$ for  gluon i=1,2 
and $\vec p_{l(k)}$ for the  quarks with label l=1,3 (k=2,4).
 The identity operator 
 ${\mathbf{1}}^{(lk)}_F$ projects onto a flavor   singlet state  
 of the created  $Q\bar Q$ pair (lk). The last term 
in Eq. (15) is the color part of the interaction  vertex with 
the    Gell-Mann matrices $\lambda^a$ acting in color space of (lk).
The color octet  wave function of the gluon i
 with polarisation vector $\vec{\epsilon}_i$
is denoted by $A^a_i$. The strength of the $g_i \rightarrow Q_l\bar Q_k$ 
transition is given by  $\gamma$. We assume that flavor symmetry is 
manifested in the quark-gluon coupling \cite{ams}, hence 
$<n\bar n|V|g> \equiv <s\bar s|V|g>=\gamma$.  
In order to  evaluate the decay amplitude $M_{G_0 \rightarrow BC}$  
we will  construct a  non-relativistic digluonium  wave function 
 from cavity QCD.
\subsection{Digluonium  wave function}
In constructing the glueball wave function we use the Bag 
Model as a guidance.  The ground state scalar glueball is realized in 
the Bag Model by  the coupling of two transverse electric (TE) modes 
of the constituent gluon $(J^{PC}=1^{+-})$ to  scalar quantum 
numbers $J^{PC}=0^{++}$ \cite{jaffe}. Using  the non-relativistic 
TE gluon mode,  proposed in Ref. \cite{iddir},  the scalar glueball 
wave function is in simple correspondence with the Bag Model solution 
and can be written    as
\begin{eqnarray}
\Psi_{G_0}(\vec r_1,\vec r_2) &= &[TE_1(1^{+-})\otimes TE_2(1^{+-})]_{J=0} \\
& = & \left\{~ \left[ \Psi_{n_1=0l_1=1}(\vec r_1)\otimes \epsilon^\mu_1\right]_{1}
\otimes  \left[ \Psi_{n_2=0l_2=1}(\vec r_2)\otimes
 \epsilon^{\mu^\prime}_2\right]_{1}~\right\}_{J=0} \Psi_C^{(12)}  \nonumber
\end{eqnarray} 
where $\Psi_{n=0l=1}(\vec r)$ is  the single  particle 
 harmonic 
oscillator wave functions as defined  
 in Appendix A.  
The spherical  components of the gluon polarisation 
vector $\vec \epsilon$ are given by $\epsilon^{\mu(\mu^\prime)}$.  
The color singlet  wave function $\Psi_C^{(12)}$  for the  two gluon  state is 
\begin{equation}
\Psi_C^{(12)}  =  \frac{1}{\sqrt{8}} \sum_{b=1}^8 A_1^b~A_2^b
\end{equation}
where we have chosen the "cartesian" basis $A^b$ with 
\begin{equation}
A^b = \frac{1}{\sqrt{2}} \sum_{ij=1}^3
 \left(\lambda_{ij}\right)^b c_i~\bar{c}_j 
\end{equation}
for the  octet representation of color $SU(3)$ with $c_i$ and  $\bar c_j$ being 
the basis states for the fundamental representations ${\bf 3}$ and $\bar{{\bf 3}}$. \\
$\Psi_{G_0}(\vec r_1,\vec r_2)$ can be recoupled  as
\begin{eqnarray}
\Psi_{G_0}(\vec r_1,\vec r_2) & = & \sqrt{3}~ \sum_{g}~ (-)^{g+1} \left\{
\begin{array}{ccc}
1 & 1& 1 \\
1 & 1& g 
\end{array} \right\}~ \sqrt{2g+1} \\
& & \cdot \left\{~\left[\Psi_{n_1=0l_1=1}(\vec r_1)\otimes \Psi_{n_2=0l_2=1}(\vec r_2)
\right]_g \otimes \left[\epsilon^\mu_1 \otimes \epsilon^{\mu^\prime}_2
\right]_g~\right\}_{J=0} \Psi_C^{(12)} . \nonumber
\end{eqnarray}
Introducing relative  and centre-of-mass 
 coordinates  $\vec r=\vec r_1-\vec r_2$ 
and $\vec{R}=1/2(\vec{r}_1+\vec{r}_2)$, respectively, 
 we perform the expansion \cite{mosh}
\begin{eqnarray}
\left[\Psi_{n_1=0l_1=1}(\vec r_1)\otimes \Psi_{n_2=0l_2=1}(\vec r_2)\right]_g 
& = & \sum_{nlNL}<nl,NL,g|n_1=0l_1=1,n_2=0l_2=1,g>  \nonumber\\
& &~~ \left[\Psi_{nl}(\vec r)\otimes \Psi_{NL}(\vec R)\right]_g 
\end{eqnarray}
where energy conservation restricts the sum to  $2n+l+2N+L=2$. The 
relevant values for the transformation brackets 
$<nl,NL,g|n_1l_1,n_2l_2,g>$ can be found in  Appendix C. \\
When  projecting out the spurious centre-of-mass excitations,  
the normalised  glueball wave function can be written in its centre-of-momentum frame as
\begin{eqnarray}
\Psi_{G_0}(\vec q_1,\vec q_2) & = & 
 \left( \sqrt{\frac{5}{9}}~\left \{~
\Psi_{n=0l=2}(\vec q_1, \vec q_2)\otimes 
\left[ \epsilon^\mu_1 \otimes \epsilon_2^{\mu^\prime} \right]_2~ 
 \right\}_0 \right.   \\
& & \left.- \frac{2}{3}~
 \left\{~\Psi_{n=1l=0}(\vec q_1, \vec q_2)\otimes \left[
\epsilon^\mu_1 \otimes \epsilon_2^{\mu^\prime} \right]_0~\right\}_0~\right) 
\delta(\vec q_1 + \vec q_2)~ \Psi_C^{(12)} \nonumber 
\end{eqnarray}
with the internal gluon momenta
$\vec q_i$ $(i=1,2)$ and $\Psi_{nl}(\vec q_1, \vec q_2)$ given in Appendix A. 
\subsection{Digluonium  decay amplitudes}
With the  digluonium wave function of Eq. (21) were are now in the 
position to evaluate the decay amplitudes for the 
process $G_0 \rightarrow BC$ of Fig. \ref{gluea}. The decay amplitude 
 $M_{G_0 \rightarrow BC}^{(l_{BC})}$, as
 defined by Eq. (3),
 separates in analogy to  quarkonia decay  into a 
flavor part $M^F_{G_0 \rightarrow BC}$ with
\begin{equation}
M^F_{G_0 \rightarrow BC}~=~ <\chi_F^B(13)\chi_F^C(42)|{\mathbf{1}}_F^{(12)}
{\mathbf{1}}_F^{(43)}|0>
\end{equation}
 and a  spin-spatial-color part
\begin{eqnarray}
M_{G_0 \rightarrow BC}^{(l_{BC})SSC} & = & 
\sqrt{2\pi K}~\sqrt{\frac{E_BE_C}{M_{f_0}}}~
M_{G_0 \rightarrow BC}^C~
 \int d\vec{q}_1 d\vec{q}_2 \prod_{i=1,..4}d\vec{p}_i~
  \delta(\vec{q}_1-\vec{p}_1-\vec{p}_2) \\
& &  \delta(\vec{q}_2-\vec{p}_4-\vec{p}_3) ~
  \Psi_C^\dagger(\vec{p}_4,\vec{p}_2)~\Psi_B^\dagger(\vec{p}_1,\vec{p}_3) 
~(\vec{\sigma}_{(12)}\cdot \vec{\epsilon_1})~
(\vec{\sigma}_{(43)}\cdot \vec{\epsilon_2})
 ~\Psi_{G_0}(\vec{q}_1,\vec{q}_2). \nonumber
\end{eqnarray}
 The flavor matrix elements in the   
 $SU(3)$ flavor symmetry limit 
are given in Table \ref{tab2}.  The 
  color transition amplitude $M_{G_0 \rightarrow BC}^C$ is given by
\begin{eqnarray}
M_{G_0 \rightarrow BC}^C & = & 
\frac{1}{4} < \chi^{(13)}_C \chi^{(42)}_C| \sum_{c=1}^8 \lambda^c_{(12)}A^c_1~
 \sum_{d=1}^8 \lambda^d_{(43)}A^d_2|\Psi_C^{(12)}>  \\
& = & \frac{1}{24\sqrt{2}}~\sum_{b=1}^8 ~Tr[(\lambda^b)^2]~=~\frac{\sqrt{2}}{3} \nonumber . 
\end{eqnarray}
The analytic evaluation of Eq. (22) contains  the 
spin matrix element 
\begin{eqnarray}
<\chi_{S_B}(13)~\chi_{S_C}(42)|
~(\vec{\sigma}_{(12)}\cdot \vec{\epsilon_1})~
(\vec{\sigma}_{(43)}\cdot \vec{\epsilon_2})|
\left[ \epsilon^\mu_1 \otimes \epsilon_2^{\mu^\prime} \right]_{S=0,2}> &  \\
=6\sqrt{(2S_B+1)(2S_C+1)}~(-)^{S_C+1}~<S_BM_BS_CM_C|S\mu+\mu^\prime>
 & \nonumber
\left\{ \begin{array}{ccc}
1/2 & 1/2 & S_B \\
1/2 & 1/2 & S_C \\
1 & 1 & S 
\end{array} \right\} & 
\end{eqnarray}
and the spatial overlap 
\newpage
\begin{eqnarray}
& &   \int d\vec{q}_1 d\vec{q}_2 \prod_{i=1,..4}d\vec{p}_i~
  \delta(\vec{q}_1-\vec{p}_1-\vec{p}_2)~ 
  \delta(\vec{q}_2-\vec{p}_4-\vec{p}_3)~ 
 \delta(\vec{q}_1+\vec{q}_2)    \\
& & \cdot \delta(\vec{p_1}+\vec p_3 -\vec{K})~  
\delta(\vec{p_4}+\vec p_2 +\vec{K})~ 
  \Psi_{n_Bl_B}^\dagger(\vec{p}_4,\vec{p}_2)~
\Psi_{n_Cl_C}^\dagger(\vec{p}_1,\vec{p}_3)~
 \Psi_{n=0,1l=2,0}(\vec{q}_1,\vec{q}_2)   \nonumber \\
& & = ~\delta_{l_f;0}~\delta_{m_f;0}~\frac{1}{8}~
\int d\vec{q}_1~d\vec{\rho}~~
\Psi_{n_Bl_B}(\vec q_1-1/2(\vec K-\vec{\rho}),-\vec q_1-1/2(\vec K+\vec{\rho}))~~
 \nonumber\\
& & ~\cdot~ \Psi_{n_Cl_C}(1/2(\vec K+\vec{\rho}),1/2(\vec K-\vec{\rho}))~~
\Psi_{n=0,1l=2,0}(\vec{q}_1,-\vec q_1)   \nonumber 
\end{eqnarray}
where $l_f$ is the relative orbital angular momentum of the final state  with 
projection $m_f$. 
A complete listing of the  analytic results for the amplitudes 
 $M_{G_0 \rightarrow BC}^{(0)SSC}$ is given  in Appendix D. \\
For the decay $G_0 \rightarrow BC~(l_f)$ 
we obtain the  dynamical selection rule that the final state 
partial wave is restricted to $l_f=0$.
Hence the decay $f_0(M) \rightarrow a_1\pi$ will 
proceed entirely by the quarkonia components of $f_0(M)$.
Moreover,  the D-wave decay amplitude $(l_f=2)$ for the two vector 
meson final state is forbidden in contrast to quarkonia decay. \\
With the $Q\bar Q$ meson size parameters fixed,
  $M_{G_0 \rightarrow BC}^{(0)SSC}$ depends  on the 
glueball size parameter $R_{G_0}$. In the constituent picture 
we expect  the mass $M_{G_0}$ of the digluonium to be 
 $M_{G_0}\approx 2~m_g$ with
 a lattice motivated constituent gluon mass of about 
 $m_g \approx 750-850~MeV$. Assuming equal confinement strength for 
quarks and gluons on the constituent level 
 we get  $R_{G_0}=3.65 /\sqrt{2.5}~GeV^{-1}$. With 
 this choice the  
constituent mass for the lightest quark flavors lies in  a
reasonable mass range of  $300-340~MeV$.  
\newpage
\section{Two-body decays of scalar-isoscalar states  and three-state mixing}
\label{clo-wein}
In the following we discuss the application of the decay models developed 
in the previous sections to the phenomenology of the $f_0$ decays. 
Given the detailed experimental decay data on the $f_0(1500)$, our analysis 
will focus on this state with decay predictions given for the partner 
states $f_0(1370)$ and $f_0(1710)$. After a brief overview 
of the experimental results for  $f_0(1500)$, we will illustrate the 
distinct decay patterns resulting from the unmixed $Q\bar Q$ 
and $G_0$ components. In a first analysis we apply the decay models to the 
three-state mixing schemes constructed in Refs. \cite{ams,weina}. Then 
we directly extract  
 the mixing coefficients from a fit 
 to the experimental two-body decay data. 
\subsection{Observed decay properties of $f_0(1500)$}
The scalar-isoscalar resonance $f_0(1500)$ is now  clearly established 
in proton-antiproton ($p\bar p$)  
 annihilation reactions  by the  Crystal Barrel 
collaboration  at CERN \cite{amsler}.
Mass and width of  $f_0(1500)$ are determined as \cite{amsler}:
\begin{equation} \label{1500}
 M_{f_0(1500)}=1505~\pm~9~MeV \qquad \Gamma_{f_0(1500)}~=~111~\pm~12~MeV .   
\end{equation}
For the two-body decay modes $f_0(1500) \rightarrow BC$ Crystal Barrel 
obtains  the branching ratios $BR(BC)$:
\begin{eqnarray}
BR(\pi\pi)  =  29\pm7.5 \% &
~~BR(\eta\eta) = 4.6\pm1.3 \% &
~~BR(\eta\eta^\prime) = 1.2\pm0.3\% \nonumber\\
BR(K\bar K)=  3.5\pm0.3\% &  
~~BR(4\pi) = 61.7\pm9.6\%&   .
\end{eqnarray}
Here, the $4\pi$ final state includes   
$\sigma\sigma$ and 
$\pi^*(1300)\pi$ as intermediate decay channels. The allowed  $\rho\rho$
decay mode of the $f_0(1500)$   
  is found to be weak \cite{thoma}.
Taking the central value for the width we deduce  from the
measured   branching ratios  
  the  partial decay widths $\Gamma(BC)$:  
\begin{eqnarray} \label{expcrystal} 
\Gamma(2\pi) =  32.2\pm8.4~MeV & 
~~\Gamma(\eta\eta)  = 5.2\pm1.5~MeV & 
~~\Gamma(\eta\eta^\prime) = 1.4\pm0.4~MeV \nonumber\\
\Gamma(K\bar K) = 3.9\pm0.4~ MeV &  
~~\Gamma(4\pi) = 68.5\pm10.7~MeV &   .
\end{eqnarray}
Alternative multi-channel
 analyses  \cite{bugga} 
 based on data sets from previous experiments have also been performed.
The results for the 
$K \bar K$ and $\eta\eta$ modes are within the experimental errors 
of the Crystal Barrel data. A larger 
total width ($\Gamma_{tot}=132\pm15~MeV$) 
and $\pi\pi$ decay width ($\Gamma_{\pi\pi}=60\pm12~MeV$) 
is obtained  suggesting a smaller  value for 
$\Gamma(4\pi)$ than indicated in Eq. (28).  
\subsection{Two-body decays of the $Q\bar Q$ and $G_0$ components}
We first discuss the decay patterns resulting 
from  the individual decay mechanisms of 
Figs. \ref{quarkonia} and \ref{gluea}, 
 coupling to the $Q\bar Q$ and the $G_0$ components of the 
$f_0$ states, respectively. \\
In  lowest order, neglecting a possible $G_0 \rightarrow G_0G_0$
 contribution, 
the decay $f_0 \rightarrow BC$ will proceed via its quarkonia 
components $Q\bar Q$. In general we consider a mixed quarkonium state
\begin{equation} \label{mixq}
|Q\bar Q>~=~cos\alpha~|n\bar n>~-~sin\alpha~|s\bar s>.
\end{equation}
The dependence of the branching ratios $BR$ for the two-body
decays of $f_0(1500)$  
 on the mixing angle $\alpha$ is given in Fig. \ref{loword}. 
The results of Fig. \ref{loword} correspond to the original discussion in Ref. \cite{ams} 
for the two pseudoscalar meson decay modes of a $^3P_0$ isoscalar 
$Q\bar Q$ state. But now predictions for all possible two-body 
decay modes are given. \\
With the three-state mixing schemes of Ref. \cite{ams} 
($a_{n\bar n}=0.43, a_{s\bar s}=-0.61$, corresponding to $\alpha=54.8^o)$ 
and Ref. \cite{weina} $(a_{n\bar n}=0.40,a_{s\bar s}=-0.90,$ that is   
 $\alpha=66^o)$ 
we obtain following  ratios for the pseudoscalar decay modes 
\begin{eqnarray}
BR(\pi\pi):BR(K\bar K): BR(\eta\eta) : BR(\eta\eta^\prime)&   & \nonumber\\
 =~1: 0.21 : 0.005 : 0.93 \qquad  \mbox{Ref. \cite{ams}} & & \\
 =~  1 : 0.97 : 0.12 : 1.80 \qquad \mbox{Ref. \cite{weina}} & &
\end{eqnarray}
to be compared with the  Crystal Barrel results \cite{amsler} of 
\begin{eqnarray}
BR(\pi\pi):BR(K\bar K): BR(\eta\eta) : BR(\eta\eta^\prime) & & \\ 
=~~ 1: 0.12 \pm 0.03 : 0.16 \pm 0.06 : 0.04 \pm 0.02 . & & \nonumber
\end{eqnarray}
Best agreement with  data is achieved for $f_0(1500) \approx |n \bar n>$,
for which we obtain 
\begin{equation}
BR(\pi\pi):BR(K\bar K): BR(\eta\eta) : BR(\eta\eta^\prime)
=1 : 0.19 : 0.16 : 0.10 .
\end{equation}
Hence, experimental data for the ratios of two pseudoscalar decay modes 
of the $f_0(1500)$ are consistent with a dominant $n\bar n$ interpretation. 
However, the predicted total width for a pure $^3P_0~n\bar n$ state 
is in conflict with experimental observation. In Fig. \ref{width}
 we indicate the calculated 
total width for a $f_0(1500)$ state with the $Q\bar Q$ configuration 
of Eq. (\ref{mixq}). 
A pure $n\bar n$ state results in $\Gamma_{tot}\approx500~MeV$, which is 
considerably higher than the observed width.
A sizable reduction of the $Q\bar Q$ amplitude in the $f_0(1500)$,  as affected 
by the admixture of a $G_0$ component,  leads to a reduction 
of the total width. For example, the mixing schemes of Refs. \cite{ams} and 
\cite{weina} lead to a total width of $\Gamma=111~MeV$ and $\Gamma=127~MeV$, 
respectively. 
Working in the lowest order of the decay mechanism, i.e. $f_0(1500)$ decays 
by its $Q\bar Q$ components, the observed two
  pseudoscalar decay modes are consistent 
with a  dominant $n\bar n$ content, slightly favoring
 the mixing scheme of Ref. \cite{ams}; the possible presence of an admixed 
$G_0$ component is then only reflected in the corresponding reduction 
of the total width. \\
When going beyond the two pseudoscalar meson decay modes
 in the leading order decay 
scheme, strong deviations from the observed decay pattern of the $f_0(1500)$ 
arise \cite{mitja}. For the two-body decay modes leading to the $4\pi$ 
decay channel,  including also charged pion combinations,  we obtain  
\begin{equation} \label{xy}
BR(\pi\pi):BR(\rho\rho):BR(\pi^*\pi):BR(\sigma\sigma):BR(a_1\pi)=
1: 0.97 :0.14 :0.31 :1.8 .
\end{equation} 
Predictions are independent of the mixing angle $\alpha$, since the 
final states of Eq. (\ref{xy}) only couple to the $n\bar n$ configuration. 
Relative to the $\pi\pi$ channel the $a_1\pi$ mode is the strongest 
decay channel. The predicted large  $\rho\rho$ branching ratio with 
$BR(\rho\rho)\approx BR(\pi\pi)$ is in  conflict with the 
preliminary analysis \cite{thoma}, where the $\rho\rho$ channel is found to 
be strongly suppressed. Furthermore, for the combined $\sigma\sigma$ and 
$\pi^\ast\pi$ decay channels with $BR(\pi^\ast \rightarrow \sigma\pi)=0.1$ 
\cite{ulrike} we deduce the ratio  
 $BR(4\pi)/BR(2\pi)=0.32$, which is much lower than the experimental
result of  $BR(4\pi)/BR(2\pi)= 2.1 \pm 0.6$ \cite{amsler}. 
The observed suppression of the $\rho\rho$
and enhancement of the 
$4\pi^0$ decay modes are both in strong conflict with a  na\"{\i}ve
 $Q\bar Q$
 interpretation  of the $f_0(1500)$ and give further indication 
for a sizable direct coupling of the gluonic component to the decay channel.\\
Next we consider the decay pattern evolving from the direct decay of the $G_0$ 
component as indicated in Fig. \ref{gluea}.
Branching ratios for the scalar  glueball $(G_0)$ decay 
 in  dependence
on its mass  are  shown  in  Fig.
\ref{gluemass}. 
A  main signature
is the strong suppression of the two vector meson decay channels, particularly 
the $\rho\rho$,  over the
entire mass range, due to
the dynamically forbidden   D-wave amplitude.
 The $\pi^\ast\pi$ and $\sigma\sigma$ decay modes exceed the contribution
 of the $\pi\pi$ decay channel 
for a $G_0$ mass of about $1500~MeV$.
These results are in clear contrast 
to those obtained for the $Q\bar Q$ decay 
mechanism. Both, the suppression of the $\rho\rho$ decay mode and the enhancement 
of the $4\pi^0$ decay channels resulting from the decay of the gluonic component 
$G_0$ are features observed in the $f_0(1500)$ decay pattern as discussed 
previously.
 For the two pseudoscalar meson  decays the results
differ from the na\"{\i}ve expectation \cite{ams}
\begin{equation} \label{gl}
BR(\pi\pi):~BR(K\bar K):~BR(\eta\eta): BR(\eta\eta^\prime)~=~1:~\frac{4}{3}
:~\frac{1}{3}:~0
\end{equation}
deduced from  flavor symmetric couplings.
 For a glueball mass of  $M_{G_0}=1500~(1700)~MeV$ we obtain 
\begin{equation}
BR(\pi\pi):~BR(K\bar K):~BR(\eta\eta)=1:~3.7~(4):~3.9~(4.3).
\end{equation}
with $BR(\eta\eta^\prime)=0$ in the SU(3)   flavor symmetry limit. 
The enhancement of the $K\bar K/\pi\pi$ decay ratio relative to 
the  na\"{\i}ve SU(3) flavor estimate of Eq. (35) is a feature 
confirmed by recent lattice results \cite{sexton}. 
For the given choice of  radial meson parameter ($R_K=2.82~GeV^{-1}, 
R_\eta=2.48~GeV^{-1}$), as extracted from tensor meson decay, we obtain 
$BR(K\bar K)\approx BR(\eta\eta)$.
 Choosing the SU(3) flavor limit of $R_K=R_\eta$ leads to
 $BR(K\bar K)=4.2~BR(\eta\eta)$,   which is again 
consistent with the analysis of Ref. \cite{sexton}.
 Hence, the proposed modelling of the $G_0$ decay mechanism
contains the main features obtained 
from  lattice calculations.\\
As a measure of flavor  symmetry breaking  we  introduce
  the  parameter $S$ defined by \\
$S\equiv <u\bar u|V|g>/<s\bar s|V|g>\not= 1$ leading to
 $BR(\eta\eta^\prime)\not= 0$. The flavor matrix elements for
$K\bar K, \eta\eta, \eta\eta^\prime$ decays  listed in Table \ref{tab1}
correspond to the exact flavor symmetry limit $(S=1)$
 and have to be generalised for pure  glueball decays $(a_{G_0}=1)$
\begin{eqnarray}
M^F_{G_0 \rightarrow K\bar K}&=& 2~\sqrt{2}~S \nonumber\\
M^F_{G_0 \rightarrow \eta\eta}&=& \sqrt{2}~(\alpha_{PS}^2
+S^2\beta_{PS}^2) \\
M^F_{G_0 \rightarrow \eta\eta^\prime}&=& 2~\alpha_{PS}~\beta_{PS}~
(1-S^2 ) . \nonumber
\end{eqnarray}
If we adopt the extreme value of $S^2=0.5$ \cite{anisovich} the  strength of
 the $K\bar K$ and $\eta\eta$ decay  modes will be
 suppressed by 50\%  and 34\%, respectively, as 
compared to the  unbroken flavor symmetry couplings.\\
For  scalar glueball masses  at $M_{G_0}= 1300,~ 1500,~ 1700~MeV$ the reduced
available phase space  suppresses  the $\eta\eta^\prime$ mode:
\begin{equation}
\begin{array}{cccccc}
M_{G_0}~[MeV] & & BR(\pi\pi):& BR(K\bar K):& BR(\eta\eta)
 :& BR (\eta\eta^\prime) \nonumber\\
1300 & & 1:& 1.3:& 2:& 0.0 \nonumber\\
1500 & & 1:&1.8:&2.6:&0.08 \\
1700 &  &
1:& 2:& 2.9:& 0.28 \nonumber .
\end{array}
\end{equation} 
Strong violation of  flavor symmetry reduces the
 $K\bar K/\pi\pi$ ratio, while leading to a
 finite $\eta\eta^\prime$ contribution. However, data
 on charmonium decay involving hard gluons indicate no significant 
symmetry breaking \cite{ams}. We therefore choose to set S=1 in the following 
discussion. \\
Decay data on $f_0(1500)$ possess qualitative 
features arising both from the decay of a $Q\bar Q$ (here dominantly 
$n\bar n$) and a $G_0$ configuration, where a pure $Q\bar Q$ or $G_0$ state 
interpretation of the $f_0(1500)$ is excluded. 
Given this simple first analysis of the decay pattern of the $f_0(1500)$ 
we are naturally led to a mixing scheme where both components 
are present.
\subsection{Decay analysis in the three-state mixing schemes}
 In the following we give a decay
analysis for the $f_0$ states defined by the three-state mixing schemes
of Refs. \cite{ams} and \cite{weina}. Now, the couplings both
of the $Q\bar Q$ and $G_0$ components to the two-meson decay
channels are included. \\
When  the physical $f_0$ mesons are given 
 as  eigenstates of the interaction Hamiltonian 
$H_I$ of  Eq. (1),  the partial decay widths  of the process
 $f_0 \rightarrow BC$  
  are  defined  by Eq. (5)  with 
the ratio of coupling strengths $\kappa$ and   the phase angle 
 $\phi$ 
 left  as free parameters. The structure of $H_I$ implies  unbroken flavor
symmetry, hence we choose $S=1$. For the $f_0(1500)$ 
decays we give in Table \ref{tab3} the  squared decay amplitudes 
and the interference terms as defined in  Eq. (5), where mass averaging 
as given by Eq. (6) is already included. 
 The mass distribution of $f_0(1500)$ is parameterised with the values of Eq. (\ref{1500}).  \\ 
The dependence of the partial decay widths  $\Gamma(BC)$ 
 on the mixing coefficients  $a_i$ $(i=n\bar n, s\bar s, G_0)$
  in the schemes of Ref. \cite{ams} ($\equiv$ scheme A) and 
Ref. \cite{weina} ($\equiv$ scheme B) are studied    by adjusting 
$\kappa$ and $\phi$ to the Crystal Barrel data
   for $f_0(1500) $ decays. 
First  we have determined $\kappa$ and $cos\phi$ from 
a fit to the observed decay modes  
 $f_0(1500) \rightarrow K\bar K$ and $\pi\pi$.  
 For both mixing schemes  we obtain 
parameter values  resulting 
in a   $ f_0 \rightarrow \eta\eta$ decay width in good agreement
with the experimental number.
This implies phenomenologically, that  
 a  $G_0 \rightarrow G_0 G_0$ contribution
to the  $\eta\eta$ decay channel can be neglected.
For this reason,  we also  take  the $\eta\eta$ decay mode into account 
when  determining $\cos\phi$ and $\kappa$  from a more restrictive fit.
 The partial decay widths   have a rather weak dependence 
 on the parameterisation
of the mass distribution of the physical $f_0$ states.  Therefore, 
the discussion \cite{amsler} on the detailed relationship
 between the  $f_0(1370)$ and 
$f_0(400-1200)$ states  is peripheral to our analysis.
For our numerical results we use the resonance parameters     
$M_{f_0(1370)}=1360~MeV,~ \Gamma_{f_0(1370)}=350~MeV$ \cite{amsler}, and 
 $M_{f_0(1710)}=1697~MeV,~ \Gamma_{f_0(1710)}=175~MeV$ \cite{pdg}.\\
The mixing coefficients in  scheme A  \cite{ams} are given as  
\begin{equation}
\left( \begin{array}{c}
|f_0(1370)> \\
|f_0(1500)>\\
|f_0(1710)>
\end{array} \right) = \left( 
\begin{array}{ccc}
0.86 & 0.13 & -0.5 \\
0.43 & -0.61 & 0.61 \\
0.22 & 0.76 & 0.6 
\end{array} \right)~ 
\left( \begin{array}{c}
|n\bar n> \\
|s\bar s> \\
|G_0> 
\end{array} \right).
\end{equation} 
From a  fit to the $f_0(1500) \rightarrow \pi\pi, K\bar K, \eta\eta$ decay 
widths  
we deduce  $cos\phi=-0.67$ and $ \kappa=0.095 ~GeV^{-2}$. Predictions  
 for the  partial decay widths of the physical $f_0$ states are  presented 
in Table \ref{tab4}. 
    For $f_0(1500)$ the  predicted 
$\eta\eta^\prime$ decay width is considerably lower than the Crystal 
Barrel value,   
 but the obtained 
 ratio $\Gamma(\eta\eta^\prime)/\Gamma(\eta\eta)$ agrees
 with the GAMS value  of $2.7\pm0.8$ for $f_0(1590)$ 
 \cite{pdg}. It is generally believed that 
$f_0(1500)$ and $f_0(1590)$ are manifestations of a single state,  
 but then the large deviation in the experimental data 
for the $\eta\eta^\prime$ decay mode has to be clarified. 
Compared to the lowest order decay analysis, where only the coupling 
to the $Q\bar Q$ component  is retained,  the  
 $\rho\rho$ mode  is more suppressed and the $4\pi$ channel enhanced.  
We obtain  $BR(\pi\pi):BR(\rho\rho):BR(4\pi)= 1:~0.65:~0.46-0.76$ for 
$BR(\pi^* \rightarrow \sigma\pi)=0.1 - 1$, 
 still in deviation from the  Crystal Barrel data. \\
In the mixing scheme B \cite{weina} the  amplitudes for the physical 
states are  given as 
\begin{equation}
\left( \begin{array}{c}
|f_0(1370)> \\
|f_0(1500)>\\
|f_0(1710)>
\end{array} \right) = \left(
\begin{array}{ccc}
0.84& 0.28 & -0.46\\
0.40 & -0.9 & 0.19 \\
0.36 & 0.34 & 0.87
\end{array} \right)~
\left( \begin{array}{c}
|n\bar n> \\
|s\bar s> \\
|G_0>
\end{array} \right).
\end{equation}
From an analogous fit we obtain  
  $cos\phi=-0.953$ and $\kappa=0.437~GeV^{-2}$. The predicted  
two-body  partial decay widths are given in Table \ref{tab5}.   
 Compared to the mixing scheme A, $f_0(1500)$
has a reduced $G_0$ component,  but 
   the experimental data demand a significant $G_0 \rightarrow 
BC$ contribution  in fitting the $\pi\pi,~K\bar K$ and $\eta\eta$ decays. 
This is balanced by giving the $G_0 \rightarrow BC$ transition 
 a   stronger weight  than in the 
mixing scheme A,  resulting in a higher  value for $\kappa$. 
Again, the predicted strength of the $\eta\eta^\prime$ 
decay mode overestimates the experimental value, while similar results 
are obtained for the $\rho\rho$ and $4\pi$ decay modes. For mixing scheme 
B we obtain 
 $BR(\pi\pi):BR(\rho\rho):BR(4\pi)= 1:~0.46:~0.55-0.97$ for
$BR(\pi^* \rightarrow \sigma\pi)=0.1 - 1$. In both 
schemes $a_1\pi$ is a dominant decay mode with $BR(\pi\pi)\approx BR(a_1\pi)$.
With  our fitting procedure, that is adjusting the relative strength 
of the $G_0$ decay contribution to the 
$\pi\pi, K\bar K$ and $\eta\eta$ decay channels of the $f_0(1500)$, 
similar results are obtained for the decay pattern of the $f_0(1500)$ 
in both mixing schemes. 
 Particular predictions for the decay modes of the partner
 states $f_0(1370)$ and $f_0(1710)$ can be
 dramatically different in the two mixing schemes. In scheme 
A the total width for $f_0(1710)$ is in good agreement with the experimental 
result of $\Gamma=175\pm9~MeV$  $(\Gamma=133\pm14~MeV)$ \cite{pdg}, 
but scheme B predicts a  unobservable 
decay width of $\Gamma \approx 2000~MeV$.   
Furthermore,  for  $f_0(1710)$ we  get 
\begin{equation}
 BR(\pi\pi)/B(K\bar K) )=\left\{ \begin{array}{ccc}
0.93 & \mbox{for} & \mbox{scheme A} \\
0.60 & \mbox{for} & \mbox{scheme B}
\end{array} \right.
\end{equation}
to be compared with the experimental value of  $0.39\pm0.14$ \cite{pdg}.\\
The two mixing schemes predict rather  similar results for the $f_0(1370)$
 state.
In  both schemes $f_0(1370)$
has  dominant $\rho\rho$ and $a_1\pi$ decay modes. The ratio 
$BR(K\bar K)/(BR(\eta\eta)$ is found to be 
\begin{equation}
BR(K\bar K)/BR(\eta\eta)=\left\{ \begin{array}{ccc}
2.6 & \mbox{for} & \mbox{scheme A} \\
1.45 & \mbox{for} & \mbox{scheme B} 
\end{array} \right.
\end{equation}
to be compared with the Crystal Barrel value of
  $ BR(K\bar K)/BR(\eta\eta) 
\approx 2.5$ \cite{amsler}. \\
The main conclusions of this section remain unchanged if we use  for 
$f_0(1500)$  a    larger $\pi\pi$ decay width 
 as reported in  Ref. \cite{bugga},  since a corresponding  
 fit leads to similar  values  for
 $\kappa$ and $cos\phi$  as for  the adjustment to the  Crystal 
Barrel data. \\
Inclusion of the direct coupling 
of the $G_0$ component of the mixed $f_0(1500)$ 
state to the two-meson decay channels gives 
an improved description of its decay features. 
For both proposed mixing schemes predictions for the 
$f_0(1500)$ decays are very similar, since the difference 
in size of the $G_0$ amplitude is compensated by adjusting 
the strength of the $G_0 \rightarrow (Q\bar Q)~(Q\bar Q)$ 
transition. \\
Also, predictions for the $f_0(1370)$ state 
do not differ too much, since both schemes assign 
a dominant $n\bar n$ component to this state. 
The key difference rests on predictions for the partner 
state $f_0(1710)$. 
 We have shown that a reduced 
$G_0$ component in $f_0(1500)$,  as proposed by Ref. \cite{weina}, 
is incompatible with the existence of a $f_0$ state at 1700 MeV as a 
narrow resonance. Conversely,  if the existence of a narrow 
  scalar component of the $f_{j=0,2}(1710)$ 
\cite{pdg}
state is  confirmed,  
 then  a non-negligible  glueball component is  
residing in the $f_0(1500)$ state. 
\newpage
\subsection{Mixing amplitudes}  
\label{analysis}
Given the shortcomings of the proposed mixing 
schemes of Refs. \cite{ams,weina}  we now directly 
extract the mixing amplitudes of the $f_0$ states in the 
given decay formalism. We proceed in an analogous fashion, requiring 
a fit to the experimental data on $f_0(1500)$ decays, while demanding
  the existence 
of a narrow resonance  $f_0(1710)$ as an 
 additional requirement.\\
  First, we start with the  unspecified mixing amplitudes 
 of   the $f_0(1500)$ state vector as defined by Eq. (2). 
The mixing amplitudes of $f_0(1500)$ are determined 
from a fit to the experimental data of  
 the $\pi\pi$,  $K\bar K$ and  $\eta\eta$ decay modes 
   with the restriction of a 
weak $\rho\rho$ decay channel.   
 Exact flavor symmetry ($S=1$) for the quark-gluon 
 coupling is assumed.
Good agreement with the experimental data requires  
$a_{n\bar n}\cdot  a_{s\bar s}<0$, i.e. a destructive 
interference between the $n\bar n$ and $s\bar s$ flavors,  and we get 
\begin{equation}
|f_0(1500)>~=~0.314~|n \bar n>~-~0.581~|s\bar s>~+~0.751~|G_0>
\end{equation}
with the ratio of coupling strengths $\kappa=0.1~GeV^{-2}$ and the phase factor 
$cos\phi=-0.92$. The  fit to the $f_0(1500)$ decays alone 
 does not fix the  phase of the $a_{G_0}$ amplitude 
relative to the quarkonia coefficients. Instead we obtain 
the condition that  
  $a_{G_0}\cdot cos\phi<0$.
For the $f_0(1500)$ state vector 
  we have taken $a_{G_0}>0$ and  $cos\phi<0$ for  reasons
 given later.  
 The relative phases between the 
state amplitudes of Eq. (43)   are 
the same as in mixing schemes A and B. 
For the  full set of 
partial decay widths we obtain the results given in Table \ref{tab6}.
 Compared to  
mixing scheme A the $\rho\rho$ mode of $f_0(1500)$ 
is suppressed by a factor of two. 
This is achieved by enhancing the $G_0$ amplitude since 
the $\rho\rho$ suppression for $f_0(1500)$ is driven by the gluonic 
component (see Sec. \ref{gluon}). An alternative  fit to 
obtain   $\rho\rho$ suppression requires   
 a strong $s\bar s$ component and a larger  value for $\kappa$ as  
resulting for example from the 
mixing scheme B. However, the existence of a narrow $f_0(1710)$ state 
 rules
 out  this 
possibility.  \\ 
For the $4\pi/2\pi$ decay branching ratio we obtain
 $BR(4\pi)/BR(2\pi)=0.57-1.04$ using 
 $BR(\pi^* \rightarrow \sigma\pi)=0.1-1$, 
 where recent results  \cite{ulrike} for 
$ \pi^\ast \rightarrow \sigma\pi$ 
favor   the lower value. This is 
still in deviation from
 the Crystal Barrel value of $BR(4\pi)/BR(2\pi)=2.1\pm0.6$ \cite{amsler}. 
 The disagreement 
possibly  hints at a   sizable 
 $G_0$ component in $\sigma$. A strong coupling of $\sigma$ to $G_0$ is 
supported  by  the experimental finding \cite{pdg}
 that $\sigma\sigma$ is the largest hadronic  branching ratio for the
charmonium  states $\chi_0$ and $\chi_2$.
 In this case,  the   
lowest order decay mechanism $G_0 \rightarrow G_0 G_0$, 
neglected here,  can enhance 
 the $\sigma\sigma$  and therefore the $4\pi$ decay mode. 
 The predicted ratio $BR(\eta\eta^\prime)/BR(\eta\eta)=2.2$ is in
 agreement with the GAMS value of  $2.7\pm0.8$ \cite{pdg},  but in conflict 
 with the  Crystal Barrel result of $0.27\pm0.10$ \cite{amsler}. 
 Breaking 
of flavor symmetry at the quark-gluon vertex, that is $S \not= 1$, 
 will further enhance the $\eta\eta^\prime$ decay width. \\
With  mass and eigenstate of   $f_0(1500)$ 
 fixed at  $1500~MeV$ and by Eq. (43), respectively, we deduce for 
  the bare  masses $m_i$ $(i=n\bar n,s\bar s, G_0)$ 
entering in $H_I$ of Eq. (1)   
\begin{eqnarray}
m_{n\bar n}=1500~MeV-3.38~z,~ & ~~m_{s\bar s}=1500~MeV+1.29~z,  \\
m_{G_0}=1500~MeV+0.18~z  & \nonumber
\end{eqnarray}
with the mixing strength $z>0$. For the bare masses 
we obtain the level ordering $m_{n\bar n}<m_{G_0}<m_{s\bar s}$, 
 which is independent of z and corresponds 
to that proposed originally  in Ref. \cite{ams}.  
When taking the alternative sign pattern of   
$a_{G_0}<0$ and $cos\phi>0$ for the fit to the 
 $f_0(1500)$ decays,   we obtain the unphysical   
  level ordering 
of the bare states with  $m_{n\bar n}>m_{s\bar s}>m_{G_0}$, 
we hence exclude this choice. 
Additionally,  for the masses $M_{<(>)}$ of the mixing
partners lying above $(M_>)$ and below $(M_<)$ $f_0(1500)$ we obtain:
\begin{eqnarray}
M_> & = & 1500~MeV~+2~z \\
M_< & = & 1500~MeV~-3.9~z. \nonumber
\end{eqnarray}
 For consistency, we call these states $f_0(1370)$ and $f_0(1710)$ and use
the values for mass and total width given
 in the  previous section for the parameterisation
 of the  respective mass distributions.
For the  mixing partners of the $f_0(1500)$  we obtain the
glueball-quarkonia content
\begin{equation}
|f_0(1710)>~=~0.149~|n\bar n>~+~0.811~|s\bar s>~+~0.565~|G_0> 
\end{equation} 
and 
\begin{equation}
|f_0(1370)>~=~0.938~|n\bar n>~+~0.070~|s\bar s>~-~0.341~|G_0>,
\end{equation}
where the amplitudes are independent
 of the particular value of z. Predictions 
 for the corresponding partial decay widths of $f_0(1710)$ and 
$f_0(1370)$ are also given in Table \ref{tab6}. 
The total width and  the ratio $BR(\pi\pi)/BR(K\bar K)=0.57$ 
of the $f_0(1710)$  are  in  good
agreement with the  experimental results of 
 of  $\Gamma=175\pm9~MeV$  $(133\pm14~MeV)$ \cite{pdg} and 
 $BR(\pi\pi)/BR(K\bar K)=0.39\pm0.14$ for $f_{j=0,2}(1710)$ \cite{pdg}. 
For the $f_0(1370)$ the predicted ratio of 
 $BR(K\bar K)/BR(\eta\eta)=2.7$
is to be compared to the experimental value $\approx 2.5$ \cite{amsler}.
 For the total width 
 we get $\Gamma_{tot}=479~MeV$, somewhat  
 larger than  the Crystal Barrel value of $\Gamma=351\pm41~MeV$
 \cite{amsler}. \\ 
For a  mixing energy  $z=43\pm31~MeV$ \cite{weinb}, 
as deduced from  lattice QCD in the quenched approximation,  we obtain 
 $m_{G_0}=1508\pm6~MeV$, 
in agreement with the lattice result
 of $m_{G_0}=1.65\pm0.15~GeV$ \cite{tep}.
 For the bare masses  we get  
$m_{n\bar n}=1355\pm105~MeV$ and $M_{s\bar s}=1556\pm40~MeV$. 
The values for the glueball-quarkonia mixing matrix element 
z favored by previous approaches are $z=77~MeV$ \cite{weina}, 
$z=64\pm13~MeV$ \cite{weinb} and $z\approx 100~MeV$, where 
 later value is deduced from the 
full nonperturbative three-state mixing scheme of Ref. \cite{ams}. 
Hence, semiphenomenological extractions of z favor a value near 
the upper limit set by the lattice result. \\
 Choosing an intermediate value of $z\approx 80~MeV$ 
we obtain for the physical masses of the mixing partners 
given by Eq. (45), $M_>=1660~MeV$ and $M_<=1190~MeV$, where 
$M_<$ is lower than the mass of the $f_0(1370)$. 
 A similar result is obtained from the full mixing approach 
of Ref. \cite{ams}, which corresponds qualitatively to our extracted scheme. \\
Given our fit to the $f_0(1500)$ decays, we obtain a 
physical mass $M_<$ and a total width which deviate from the corresponding 
data on $f_0(1370)$. Therefore, within the framework of a
three-state mixing scheme, the lower lying partner of $f_0(1500)$ could 
be identified with the very broad structure around 1100 MeV \cite{ams}, 
 called $f_0(400-1200)$ \cite{pdg}. Alternatively, the Crystal Barrel 
state $f_0(1370)$ could be the high mass tail of $f_0(400-1200)$. 
 However, given the ansatz of Eq. (1), the most stable and testable 
consequences of our mixing scheme concern predictions for all the 
two-body decay modes of the $f_0$ states as given in Table 
\ref{tab6}. 
  Given the sensitivity of the decay pattern on 
the particular size of the $G_0$ and $Q\bar Q$ components 
in the $f_0$ states only, a full experimental determination 
 provides a stringent test.
\newpage
\section{Summary and conclusions}
\label{summary}
We have performed a detailed study of the two-body  decay properties of the 
 scalar-isoscalar  $f_0(1370),~f_0(1500)$ 
 and $f_0(1710)$  states resulting from the mixture of the lowest 
lying scalar glueball with  the isoscalar quarkonia states of the $0^{++}$ 
nonet. In the decay analysis we have taken into account the coupling 
of the quarkonia and glueball components of the $f_0$ states to the quarkonia 
components of the two-meson final state, where the decay 
dynamics is guided by strong coupling QCD. 
 Leading order corresponds to the transitions $Q\bar Q \rightarrow 
(Q\bar Q)~(Q\bar Q)$ and $G_0 \rightarrow G_0~G_0$, where the
latter transition, which can contribute to $\eta\eta,~\eta\eta^\prime$ and 
$\sigma\sigma$ final states, is omitted due to its ill-constrained 
nature. 
The coupling of the quarkonia component $(Q\bar Q)$ of the $f_0$
 states  is described in the framework of the $^3P_0$ 
pair creation model, which is successful in the phenomenology 
of OZI-allowed meson decay. 
In next-to-leading order we obtain a direct coupling of the $G_0$ 
component of the $f_0$ states to the quarkonia component of the  
decay channel. The corresponding transition
 $G_0 \rightarrow (Q\bar Q)~(Q\bar Q)$ 
is modelled by resorting to a scalar digluonium wavefunction for $G_0$ 
as given by cavity QCD. Predictions for the two-pseudoscalar decay 
channels, that is $\pi\pi,~K\bar K$ and $\eta\eta$, are in rough agreement 
with recent  lattice results in the limit of unbroken flavor SU(3). \\
Our analysis on the two-body decay modes of the $f_0$ states focuses 
on the $f_0(1500)$, where detailed experimental data are   
available. In comparison to the observed decay properties of 
$f_0(1500)$ we elaborate the distinct decay patterns 
resulting  from the unmixed 
$Q\bar Q$ and $G_0$ components. Taking into account the coupling of the 
$Q\bar Q$ components of the $f_0(1500)$ only, which corresponds to the 
leading order decay mechanism of Ref. \cite{ams}, data on two-pseudoscalar 
decay channels are consistent with a dominant $n\bar n$ structure 
 residing in this state. 
However, $f_0(1500)$ is too narrow for  a simple $n\bar n$ structure,  
 implying a sizable  reduction of 
 its quarkonia components as  resulting 
from glueball-quarkonia mixing.    
 When going beyond the two-pseudoscalar decay channels large deviations 
from the observed decay pattern of the $f_0(1500)$ arise.
The large $4\pi$ branching ratio, 
fed by the $\sigma\sigma$ and $\pi^\ast \pi$ 
decay channels,   and the  suppression  of  $\rho\rho$
is  in clear conflict with the  simple  leading
 order analysis of $f_0(1500)$, where coupling to its $Q\bar Q$ 
component is taken,   
 and
points towards the relevance of a possible  
 glueball component.    
      Coupling of the $G_0$ component of the $f_0(1500)$ to the 
quarkonia component of the two-body final state yields a significantly 
different decay pattern. In the two-pseudoscalar sector strong 
$K\bar K$ and $\eta\eta$ decay modes are found. Furthermore, 
$\pi^\ast \pi$ and $\sigma\sigma$ dominate over the $\pi\pi$ decay channel, 
whereas $\rho\rho$ is strongly suppressed. This latter feature 
of a strongly reduced $\rho\rho$ coupling to $G_0$
 is also obtained in a recent  work \cite{jin}, where the coupling of 
$G_0$ to the pionic decay channel is set up in an effective 
linear sigma model. Both observed features of the $f_0(1500)$, the enhancement 
of the $4\pi^0$ decay channels and the reduction of the $\rho\rho$ 
channel, are driven by its gluonic component.\\ 
In a next step we have investigated the decay properties of the $f_0(1370), ~
f_0(1500)$ and $f_0(1710)$ states in the proposed three-state 
mixing schemes of Refs. 
\cite{ams,weina}. Due to the inclusion of the  
$G_0 \rightarrow (Q\bar Q)~(Q\bar Q)$  mechanism 
in the decay analysis we were able 
to discriminate between the mixing schemes, in particular 
with respect to  
 the orthogonal physical consequences 
for  the $f_0(1710)$ state.  
 A  negligible 
$G_0$ component in $f_0(1500)$ as proposed by Ref. \cite{weina} is
 incompatible  with the existence of a narrow $f_0$ state
 in the 1700 MeV mass region; 
hence our first analysis of the two-body decays favors the mixing scheme 
of Ref. \cite{ams}. \\ 
Finally, we have extracted a three-state mixing scheme based on a detailed 
fit to the experimental $f_0(1500)$ decays, while demanding 
the existence of a narrow $f_0$ mixing partner in the 1700 MeV 
mass region. 
For the bare masses before mixing we obtain the level ordering 
$m_{n\bar n}<m_{s\bar s}<m_{G_0}$, corresponding qualitatively 
to the scheme originally proposed in Ref. \cite{ams}. 
Adjustment of the gluonic decay mechanism leads to a proper 
description of the observed two-pseudoscalar decay modes of $f_0(1500)$ 
with the exception of $\eta\eta^\prime$. Accordingly, we obtain a significant 
$\rho\rho$ suppression coupled at the same time to the enhancement 
of the $4\pi^0/2\pi$ ratio, in line with experimental 
requirements. A full reduction of the observed $4\pi/2\pi$ ratio 
requires a strong $\sigma\sigma$ mode. 
Both deviations in the $\eta\eta^\prime$ and $\sigma\sigma$ decay channels 
suggest a non-negligible leading order $G_0 \rightarrow G_0~G_0$
 contribution, which was neglected here. 
A key signature for the extracted state composition of $f_0(1500)$ is the 
prediction for the $a_1\pi$ decay mode with $BR(a_1\pi) \approx BR(\pi\pi)$. 
A pure $Q\bar Q$ configuration or equivalently, application 
of the leading order decay mechanism to the mixed $f_0(1500)$ state results 
in  $BR(a_1\pi) \approx~2  BR(\pi\pi)$. Hence an experimental 
determination of the $a_1\pi$ decay channel of $f_0(1500)$ is desirable 
to further quantify the relevance of its gluonic component. \\
The resulting predictions for the decays of the $f_0(1370)$ and 
$f_0(1710)$ states are consistent with the limited experimental 
data  \cite{amsler,pdg} 
on pseudoscalar decay modes. Clearly a search for the decay 
channels feeding the $4\pi$ decay products is most relevant in 
clarifying the nature of these partner states. Predictions for the 
total widths of $f_0(1370)$ and $f_0(1710)$ roughly agree with the observed 
values \cite{amsler,pdg}; 
the physical mass spectrum resulting from the 
extracted mixing scheme points towards a sizable influence of the broad 
 $f_0$ structure  at 1100 MeV \cite{amsler}, which should be clarified. \\
 The quantitative predictions of our analysis emphasise the role of the 
gluonic components in the $f_0$ states and its observable consequences  
in the two-body decay modes. Hence a full determination of the hadronic 
decay properties provides a sensitive test on the nature 
of the $f_0$ states, revealing the intrusion of the glueball ground state 
in the scalar-isoscalar meson spectrum.  
\begin{acknowledgments}
This work was supported in part by the Graduiertenkolleg "Struktur und
Wechselwirkung von Hadronen und Kernen" (DFG Mu705/3) and  by a grant of
the Deutsches Bundesministerium
f\"ur Bildung und Forschung (contract No. 06 T\"u 887). 
\end{acknowledgments}

\newpage
\begin{appendix}
\section{Spatial wave functions}
For the spatial part of the wave functions 
used in the text we use   harmonic oscillators
 $\Psi_{nl}(\vec p_i, \vec p_j)$ which are defined  
in the limit of equal constituent  masses  as
\begin{eqnarray}
\Psi_{nl}(\vec{p_i},\vec{p_j})&=& N_{nl}~\left(\frac{R}{2}\right)^l~
|\vec{p_i}-\vec{p_j}|^l~
\exp\left\{ - \frac{1}{8} R^2 ( {\vec{p}}_i-{\vec{p}}_j)^2 \right\} \nonumber\\
& & L_n^{l+1/2}(\frac{R^2}{4}(\vec{p_i}-\vec{p}_j)^2)~Y_{lm}(\hat{\vec p_i-\vec p_j}) 
\end{eqnarray}
with the internal  momentum $p_{i(j)}$ and the size parameter 
$R$. The normalisation constant $N_{nl}$ depends on the radial  and orbital 
 quantum numbers n, l with
\begin{equation}
N_{nl}=\left[~ \frac{2(n!)R^3}{\Gamma(n+l+3/2)}\right]^{1/2}.
\end{equation}
 The Laguerre polynomials are given by
\begin{equation}
L_n^{l+1/2}(p)=\sum_{k=0}^{n}\frac{(-)^k~
\Gamma(n+l+3/2)}{k!~\Gamma(k+l+3/2)}~p^k.
\end{equation}
For the $ Q\bar Q$  mesons we use: 
\begin{equation}
(nl) = \left\{ \begin{array}{ccl}
(00)&\mbox{for}&\pi,~\eta,~\eta^\prime,K,~\bar K,~\rho,~\omega \\ 
 (01)&\mbox{for}&a_1,~\sigma, ~f_0  \\
 (10)&\mbox{for}&\pi^*(1300) \end{array}\right. 
\end{equation}
\newpage
\section{Quarkonia decay amplitudes} 
In the following we give the full expressions for the 
 spin-spatial-color amplitudes  $M^{(l_{BC})SSC}_{Q\bar Q \rightarrow BC}$
 defined in Eq. (12) as obtained for the quarkonia decay process 
$Q\bar Q \rightarrow BC$.  
For  equal size parameters  ($\equiv R_B$)
 of  the final state 
mesons B and C we obtain: \\ 
1) $Q \bar Q \rightarrow \pi\pi,~\eta\eta,~\eta\eta^\prime,~K\bar K$\\
\begin{eqnarray}
M^{(0)SSC}_{Q\bar Q \rightarrow BC}& = & \sqrt{\frac{E_BE_C}{M_{f_0}}}~
\frac{8}{\pi^{1/4}}~\frac{R_{Q\bar Q}^{5/2}R_B^3}{(R_{Q\bar Q}^2+2R_B^2)^{5/2}}~
\exp\left\{ -\frac{1}{4}\left(
\frac{R_{Q\bar Q}^2R_B^2}{R_{Q\bar Q}^2+2R_B^2}\right)~K^2\right\} \\
& & \left[ K^{1/2}-\frac{R_B^2(R_{Q\bar Q}^2+R_B^2)}{3(R_{Q\bar Q}^2+2R_B^2)}~
K^{5/2}\right]\nonumber
\end{eqnarray}
2) $Q\bar Q \rightarrow \rho\rho,~\omega\omega$\\
\begin{eqnarray}
M^{(l_{BC})SSC}_{Q\bar Q \rightarrow BC}& = & \sqrt{\frac{E_BE_C}{M_{f_0}}}~
\frac{8}{\sqrt{3}\pi^{1/4}}~\frac{R_{Q\bar Q}^{5/2}R_B^3}{(R_{Q\bar Q}^2+2R_B^2)^{5/2}}~
\exp\left\{ -\frac{1}{4}\left(\frac{R_{Q\bar Q}^2R_B^2}{R_{Q\bar Q}^2+2R_B^2}\right)~K^2\right\} \\
& & \cdot \left\{ \begin{array}{cc}
    K^{1/2}-\frac{R_B^2~(R_{Q\bar Q}^2+R_B^2)}{3~(R_{Q\bar Q}^2+2R_B^2)}~K^{5/2} & \mbox{for} 
\qquad
l_{BC}=0 \\
  & \\
-~\sqrt{\frac{8}{9}}~\cdot~ \frac{R_B^2(R_{Q\bar Q}^2+R_B^2)}{R_{Q\bar Q}^2+2R_B^2}~K^{5/2}  & \mbox{for} \qquad l_{BC}=2 
\end{array} \right.
\end{eqnarray}
3) $Q\bar Q \rightarrow \sigma\sigma$ \\
\begin{eqnarray}
M^{(0)SSC}_{Q\bar Q \rightarrow \sigma\sigma}& = & \sqrt{\frac{E_\sigma E_\sigma}{M_{f_0}}}~
\frac{4}{9\pi^{1/4}}~\frac{R_{Q\bar Q}^{5/2}R_B^5}{(R_{Q\bar Q}^2+2R_B^2)^{7/2}}~
\exp\left\{ -\frac{1}{4}\left(
\frac{R_{Q\bar Q}^2R_B^2}{R_{Q\bar Q}^2+2R_B^2}\right)~K^2\right\} \\
& &\left[60K^{1/2}+\frac{11R_{Q\bar Q}^4+4R_B^2(R_{Q\bar Q}^2+R_B^2)}{R_{Q\bar Q}^2+2R_B^2}~K^{5/2}
-\frac{R_{Q\bar Q}^4R_B^2(R_{Q\bar Q}^2+R_B^2)}{(R_{Q\bar Q}^2+2R_B^2)^2}~K^{9/2}\right]
\nonumber
\end{eqnarray}
4) $Q\bar Q \rightarrow \pi^*(1300)\pi$\\
\begin{eqnarray}
M^{(0)SSC}_{Q\bar Q \rightarrow \pi^*(1300)\pi}& = & \sqrt{\frac{E_{\pi}E_{\pi^*}}{M_{f_0}}}~
\frac{2^{3/2}}{3^{3/2}\pi^{1/4}}~\frac{R_{Q\bar Q}^{5/2}R_B^3}{(R_{Q\bar Q}^2+2R_B^2)^{7/2}}~\exp\left\{ -\frac{1}{4}\left(
\frac{R_{Q\bar Q}^2R_B^2}{R_{Q\bar Q}^2+2R_B^2}\right)~K^2\right\} \nonumber\\
& & \left[ 6\cdot (3R_{Q\bar Q}^2-4R_B^2)~K^{1/2}-\frac{R_{Q\bar Q}^2R_B^2(13R_{Q\bar Q}^2
+6R_B^2)}{R_{Q\bar Q}^2+2R_B^2}~K^{5/2} \right. \\
& & \left.~+~
\frac{R_{Q\bar Q}^4R_B^4(R_{Q\bar Q}^2
+R_B^2)}{(R_{Q\bar Q}^2+2R_B^2)^2}~K^{9/2}\right] \nonumber
\end{eqnarray}
5) $Q\bar Q \rightarrow a_1\pi$\\
\begin{eqnarray}
M^{(1)SSC}_{Q\bar Q \rightarrow a_1\pi}& = &  \sqrt{\frac{E_{a_1}E_\pi}{M_{f_0}}}~
 \frac{2^4}{3\pi^{1/4}}
\frac{R_{f_0}^{5/2}R_B^4}{(R_{Q\bar Q}^2+2R_B^2)^{5/2}}
\exp\left\{ -\frac{1}{4}\left(
\frac{R_{Q\bar Q}^2R_B^2}{R_{Q\bar Q}^2+2R_B^2}\right)~K^2\right\}K^{3/2}
\end{eqnarray}
\newpage
\section{Transformation brackets}
The relevant values for the transformation brackets 
 $<nl,NL,g|n_1l_1,n_2l_2,g>$ 
used in Eq. (19) are: \\ 
1) $g=0$ 
\begin{eqnarray}
<00,10,0|01,01,0> & = & \frac{1}{\sqrt{2}} \\
<01,01,0|01,01,0> & = & 0 \\
<10,00,0|01,01,0> & = & -\frac{1}{\sqrt{2}} 
\end{eqnarray}
2) $g=1$ 
\begin{eqnarray}
<01,01,1|01,01,1> & = & 1 
\end{eqnarray}
3) $g=2$ 
\begin{eqnarray}
<00,02,2|01,01,2> & = &  \frac{1}{\sqrt{2}} \\
<01,01,2|01,01,2> & = & 0 \\
<02,00,2|01,01,2> & = &  -  \frac{1}{\sqrt{2}} 
\end{eqnarray}
\newpage
\section{Digluonium decay amplitudes}
The spin-spatial-color matrix elements $M^{SSC}_{G_0 \rightarrow BC}$, 
defined in Eq. (22), 
 for the decay of the scalar digluonium state $G_0$ (size parameter $R_{G_0}$)
 into the final state mesons  B C (size parameter $R_B$) are: \\
1) $G_0 \rightarrow \pi\pi,~\eta\eta,~\eta\eta^\prime,~K\bar K~( \equiv 2~PS)$
\begin{eqnarray}
M^{(0)SSC}_{G_0 \rightarrow 2PS} & = & \sqrt{\frac{E_BE_C}{M_{f_0}}}~
\frac{32}{3}~\sqrt{2}~\pi^{7/4}~\frac{R_{G_0}^{3/2}~(R_B^2-2R_{G_0}^2)}{(R_B^2+2R_{G_0}^2)^{5/2}}~K^{1/2}
\end{eqnarray} 
2) $G_0 \rightarrow \rho\rho,~\omega\omega~( \equiv 2~VM)$
\begin{eqnarray}
M^{(0)SSC}_{G_0 \rightarrow 2VM} & = & \frac{1}{\sqrt{3}}~
M^{(0)SSC}_{G_0 \rightarrow 2PS}
\end{eqnarray}
3) $G_0 \rightarrow \sigma\sigma$
\begin{eqnarray}
M^{(0)SSC}_{G_0 \rightarrow \sigma\sigma} & = 
& \sqrt{\frac{E_\sigma E_\sigma}{M_{f_0}}}~
\frac{64\sqrt{2}}{9}~\pi^{7/4}~
\frac{R_{G_0}^{7/2}~(2R_{G_0}^2-9R_B^2)}{(R_B^2+2R_{G_0}^2)^{7/2}}~K^{1/2}
\end{eqnarray}
4) $G_0 \rightarrow \pi^*(1300)\pi$
\begin{eqnarray}
M^{(0)SSC}_{G_0 \rightarrow \pi^*\pi} & = & \sqrt{\frac{E_\pi^* E_\pi}{M_{f_0}}}~
\frac{32}{3\sqrt{3}}~\pi^{7/4}~
\frac{R_{G_0}^{3/2}~R_B^2~(14R_{G_0}^2-3R_B^2)}{(R_B^2+2R_{G_0}^2)^{7/2}}~K^{1/2}
\end{eqnarray} 
\end{appendix}
\newpage

\newpage
\begin{figure}
\caption{\label{glueb}Leading order decay mechanism of
the $G_0 \rightarrow G_0 G_0$ transition. The dot indicates the three-gluon
coupling.}
\end{figure}
\begin{figure}
\caption{\label{quarkonia}Quark line diagram for the
  decay of the quarkonia
component $Q\bar Q$ into the final state mesons BC occurring
 as a leading order decay mechanism in strong coupling QCD.
 The dot indicates the $^3P_0$ vertex with  strength
$\lambda$.}
\end{figure}
\begin{figure}
\caption{\label{gluea}Transition $G_0 \rightarrow BC$ occurring 
as a next-to-leading order decay mechanism 
in strong coupling QCD. The dots indicate
the quark-gluon coupling with strength $\gamma$.}
\end{figure}
\begin{figure}
\caption{\label{loword}
Dependence of the branching ratios   $BR(f_0(1500) \rightarrow BC)$ on
the mixing angle  $\alpha$,  defined by the $Q\bar Q$ configuration
 $|f_0(1500)>=cos\alpha |n\bar n>-sin\alpha |s\bar s>$, 
 for the decay mechanism $Q\bar Q \rightarrow BC$ of Fig. 2. 
 For the meson size
parameters we take the values given in the main text. The arrows indicate
the predictions for the mixing schemes  of Refs. \protect\cite{ams,weina}
  in lowest order of  the decay mechanism.}  
\end{figure}
\begin{figure}
\caption{\label{width}Dependence of the total width of the $f_0(1500)$ 
resonance on the mixing angle
$\alpha$ defined by 
the $Q\bar Q$ configuration 
 $|f_0(1500)>=cos\alpha |n\bar n>-sin\alpha |s\bar s>$.
 The shadowed band  corresponds to  the experimental total width of
 $\Gamma=111\pm12~MeV$ \protect\cite{amsler}.}
\end{figure}
\begin{figure}
\caption{\label{gluemass}
Dependence of the branching ratios $BR(G_0 \rightarrow BC)$
 on the glueball mass for the decay mechanism 
 $G_0 \rightarrow BC$ of Fig. 3.
Exact flavor $SU(3)$ symmetry  for the quark-gluon vertex  is assumed.
 For the size parameters we use 
  the values given in the main text. For mass averaging we take
 a  glueball  width of $\Gamma_{G_0}=120~MeV$.
}
\end{figure}
\newpage
\begin{table}
\caption{ Flavor matrix elements for  the two-body 
 decay $Q\bar Q \rightarrow BC$  of the mixed $f_0(M)$ states 
as defined in Eq. (11). The quarkonia
 component of the $f_0$ states is defined as 
 $|Q\bar Q>=a_{n\bar n} |n\bar n>+a_{s\bar s}|s\bar s>$.
The quantities  $\alpha_{PS}$ and $\beta_{PS}$ are related to the 
pseudoscalar mixing angle as defined 
by Eq. (13).
  For the  vector mesons we use  ideal mixing.} 
\begin{tabular}{lcl}  
{\Large B~C} & & {\Large $M_{Q\bar Q \rightarrow BC}^F$}\\
 \hline
$ \pi\pi,\rho\rho$ & & $\sqrt{3}~a_{n\bar n}$ \\
$ \sigma\sigma,\omega\omega$ & & $a_{n\bar n}$ \\
$ \pi^*(1300)\pi,a_1\pi$ & & $\sqrt{6}~a_{n\bar n}$ \\
$ K\bar K$ & & $ a_{n\bar n}+\sqrt{2}~a_{s\bar s}$ \\
$ \eta\eta$ & & $\alpha_{PS}^2~a_{n\bar n}~+~\sqrt{2}~
\beta_{PS}^2~a_{s\bar s}$ \\
$\eta\eta^\prime$ & & $2~\alpha_{PS}~\beta_{PS}~(a_{n\bar n}
/\sqrt{2}-a_{s\bar s})$
\end{tabular}
\label{tab1}
\end{table}
\newpage
\begin{table}
\caption{ Flavor matrix elements  of Eq. (21) 
 for the two-body decay $G_0 \rightarrow BC$  
of the mixed $f_0(M)$ states  
with   glueball component $a_{G_0} |G_0>$.
The quark-gluon coupling is assumed to be flavor independent.}
\begin{tabular}{lcl}  
{\Large B~C} & & {\Large $M_{G_0 \rightarrow BC}^F/a_{G_0}$}\\
 \hline
$ \pi\pi,\rho\rho$ & & $\sqrt{6}$ \\
$ \sigma\sigma,\omega\omega$ & & $\sqrt{2}$ \\
$ \pi^*(1300)\pi,a_1\pi$ & & $\sqrt{12}$ \\
$ K\bar K$ & & $ 2~\sqrt{2}$ \\
$ \eta\eta$ & & $ \sqrt{2}$ \\
$ \eta\eta{^\prime} $ &  &  0
\end{tabular}
\label{tab2}
\end{table}
\newpage
\begin{table}
\caption{ Squared decay amplitudes and interference terms 
as defined by Eq. (5) 
 for the transition $f_0(1500) \rightarrow BC$.  
Mass averaging as in Eq. (6) is already included. 
For the interference term we have  $l_{BC}=0$  
as restricted by the dynamical selection rule of the $G_0 \rightarrow BC$ 
transition. 
 All values are given in MeV.}
\begin{tabular}{llll} 
{\Large BC}& {\Large  $M_{Q\bar Q \rightarrow BC}^2$}
& {\Large $ M_{G_0 \rightarrow BC}^2$} &
{\Large $ M_{G_0 \rightarrow BC}^{(0)}
 M_{Q\bar Q \rightarrow BC}^{(0)}$} \\
\hline
$\pi\pi$ & $52.2a_{n\bar n}^2$ & $224.4a_{G_0}^2$ & $-103.1~a_{n\bar n}
a_{G_0}$ \\
$\eta\eta$ & $20.9 (0.63a_{n\bar n}+0.523a_{s\bar s})^2$&
$ 865.8 b_{G_0}^2$ & $-131.1a_{G_0} (0.63a_{n\bar n}+0.523a_{s\bar s})$ \\
$\eta\eta^\prime$ & $10.8(a_{n\bar n}/\sqrt{2}-a_{s\bar s})^2$ & 0 & 0 \\
$ K\bar K$ &$ 10.7 (a_{n\bar n}+\sqrt{2}a_{s\bar s})^2$ & $828.0 a_{G_0}^2$ &
$-86.8 a_{G_0}(a_{n\bar n}+\sqrt{2}a_{s\bar s})$ \\
$\rho\rho$ & $50.4 a_{n\bar n}^2$ & $15.0 a_{G_0}^2$ &
 $10.2 a_{n\bar n} a_{G_0}$ \\
$\omega\omega$ & $5.3 a_{n\bar n}^2$ & $ 2.2 a_{G_0}^2$& $2.0  a_{n\bar n} a_{G_0}$ \\
$a_1\pi$ & $94.2  a_{n\bar n}^2$ & 0 & 0 \\
$\sigma\sigma$ & $16.1  a_{n\bar n}^2$ & $264.6  a_{G_0}^2$
& $-63.1 a_{n\bar n} a_{G_0}$ \\
$\pi^*(1300)\pi$ & $7.2  a_{n\bar n}^2$ & $440.4  a_{G_0}^2$& $-55.2
  a_{n\bar n} a_{G_0}$
\end{tabular}
\label{tab3}
\end{table}
\newpage
\begin{table}
\caption{ Predictions of mixing scheme A \protect\cite{ams} for the 
 partial decay widths $\Gamma(BC)$  of the 
physical $f_0$ states, B C are the final state
mesons. The $4\pi$ decay channel  
refers to the contributions of the $\sigma\sigma$ and $\pi^\ast\pi$ 
intermediate states using  
 $BR(\pi^\ast \rightarrow \sigma 
\pi)=0.1-1$. All values are given in MeV.
 The predicted ratios
 $BR(f_0(1370) \rightarrow K\bar K)/BR(f_0(1370) \rightarrow \eta\eta)=2.6$ 
and $BR(f_0(1710) \rightarrow \pi\pi)/BR(f_0(1710) \rightarrow K\bar K)=0.93$ 
 have to be compared to the experimental values of
 $\approx 2.5$ \protect\cite{amsler} and $0.39\pm0.14$
 \protect\cite{pdg}.} 
\begin{tabular}{lccccccccccc}  
&$\Gamma_{tot}$  & $\pi\pi$ & $K\bar K$ & $\eta\eta$&
$\eta\eta^\prime$& $\rho\rho$ &
 $ \omega\omega$ & $a_1\pi$ &
$\pi^*\pi$&$ \sigma\sigma$
& $4\pi$   \\ \hline
&&     & \multicolumn{4}{c}{{\large $f_0(1500)$}} &  & & & \\
Model &136  & 29.3&  3.9& 5.2 &
19 &
19 & 2 & 36.8 &
9.9 &
12.5& 13.5-22.4  \\  
Exp.\cite{amsler} &  $111\pm12$  & $32\pm8.4$ & $3.9\pm0.3$ & $5.2\pm1.5$ & 
$1.4\pm0.4$&  & & & & & $68\pm10.7$  \\ \hline
&&    &  \multicolumn{4}{c}{{\large $f_0(1370)$}} &  & & & \\
Model & 410&   65.4 & 29.1& 10.8 & 3 & 112.1 & 22.8 & 138.6& 
  12.7 & 16 & 17.3-28.7  \\
Exp.\cite{amsler} &  $351\pm41$ & &\multicolumn{2}{c}{$K\bar K/\eta\eta \approx 2.5$} &&&&&& \\ \hline
& & &   \multicolumn{4}{c}{{\large $f_0(1710)$}}&   & & & \\
Model & 179 &  23.4 & 25.1 & 22.6 & 20.3 & 21.3 &
 4.3 & 28.2 & 20.9 & 12.7 & 14.8-33.6  \\
Exp.\cite{pdg} &   $133\pm14$ &\multicolumn{3}{c}{$\pi\pi/K\bar K=0.39\pm0.14$}&&&&&& 
\end{tabular}
\label{tab4}
\end{table}
\newpage
\begin{table}
\caption{ Predictions of mixing scheme B \protect\cite{weina} for the
 partial decay widths $\Gamma(BC)$  of the 
physical $f_0$ states. 
 The $4\pi$ decay channel                                       
refers to the contributions of the 
$\sigma\sigma$ and $\pi^\ast\pi$
intermediate states using
 $BR(\pi^\ast \rightarrow \sigma                                              
\pi)=0.1-1$. All values are given in MeV.
 The  predicted ratios   $BR(f_0(1370) \rightarrow K\bar K)/BR(f_0(1370)
 \rightarrow \eta\eta)=1.45$
and $BR(f_0(1710) \rightarrow \pi\pi)/BR(f_0(1710) \rightarrow K\bar K)=0.60$
have to be compared to the experimental values of
$\approx 2.5$ \protect\cite{amsler} and $0.39\pm0.14$ \protect\cite{pdg}.} 
\begin{tabular}{lccccccccccc}  
&$\Gamma_{tot}$  & $\pi\pi$ & $K\bar K$ & $\eta\eta$&
$\eta\eta^\prime$& $\rho\rho$ &
 $ \omega\omega$ &  $a_1\pi$ &
$\pi^*\pi$& $ \sigma\sigma$
& $4\pi$ \\ \hline
& &  & &  \multicolumn{4}{c}{{\large $f_0(1500)$}}   & & & \\
Model &157&   34.7&  3.9& 5.1 &
31.8 &
15.8 & 1.6 & 31.8 &
16.2 &
17.4& 19-33.6 \\ 
Exp. \cite{amsler} &  $111\pm12$  & $32\pm8.4$ & $3.9\pm0.3$ & $5.2\pm1.5$ &
$1.4\pm0.4$&  & & & & & $68\pm10.7$  \\ \hline
 & &&& \multicolumn{4}{c}{{\large $f_0(1370)$}}   & & & \\
Model &370 &  38.7 & 31.4& 21.6 & 1.3 & 110.8 & 22.6 & 132.2&
 8.2 & 2.8 & 3.6-11 \\
Exp. \cite{amsler} &  $351\pm41$ & &\multicolumn{2}{c}{$K\bar K/\eta\eta \approx 2.5$}&&&&&& \\ \hline
& &  & &\multicolumn{4}{c}{{\large $f_0(1710)$}}   & & & \\
Model &2000  & 253 & 381 & 442 & 0.4 & 63 & 12& 76& 536  & 263 &316-799 \\
Exp. \cite{pdg} &   $133\pm14$ &\multicolumn{3}{c}{$\pi\pi/K\bar K=0.39\pm0.14$}&&&&&& 
\end{tabular}
\label{tab5}
\end{table}
\newpage
\begin{table}
\caption{ Predictions  for the
 partial decay widths $\Gamma(BC)$  of the
physical $f_0$ states for the extracted mixing scheme. 
 The $4\pi$ decay channel                                       
refers to the contributions of the $\sigma\sigma$ and $\pi^\ast\pi$
intermediate states using
$BR(\pi^\ast \rightarrow \sigma                                              
\pi)=0.1-1$.
 All values are given  in MeV. 
 The  predicted ratios  $BR(f_0(1370) \rightarrow K\bar K)
/BR(f_0(1370) \rightarrow \eta\eta)=2.7$
and $BR(f_0(1710) \rightarrow \pi\pi)/BR(f_0(1710) \rightarrow K\bar K)=0.57$
 have to be compared to the experimental values of
 $\approx 2.5$ \protect\cite{amsler} and $0.39\pm0.14$ \protect\cite{pdg}.}
\begin{tabular}{lccccccccccc}  
&$\Gamma_{tot}$ & $\pi\pi$ & $K\bar K$ & $\eta\eta$&
$\eta\eta^\prime$& $\rho\rho$ &
 $ \omega\omega$ &$a_1\pi$&
$\pi^*\pi$&$ \sigma\sigma$
& $4\pi$ \\ \hline
&  && &  \multicolumn{4}{c}{{\large $f_0(1500)$}}   & & & \\
Model & 101&  23&  2.8& 6.7 &
14.6 &
9.7 & 1 & 19.6 &
11.8 &
12.1& 13.3-23.9 \\ 
Exp. \cite{amsler} &  $111\pm12$  & $32\pm8.4$ & $3.9\pm0.3$ & $5.2\pm1.5$ &
$1.4\pm0.4$&  & & & & & $68\pm10.7$  \\ \hline
 && &&\multicolumn{4}{c}{{\large $f_0(1370)$}}   & & & \\
Model &479 & 78.2 & 27.4& 10.1 & 4.6 & 133.2 & 27.1&164.9 
 & 14.6 & 19 & 20.5-33.6 \\ 
Exp. \cite{amsler} &  $351\pm41$ & &\multicolumn{2}{c}{$K\bar K/\eta\eta \approx
 2.5$}&&&&&& \\ \hline
& && & \multicolumn{4}{c}{{\large $f_0(1710)$}}   & & & \\
Model & 146 & 15 & 26.3 & 24.5 & 27.7 & 9.7 & 1.9&13&  18  & 10.1 &11.9-28.1 \\
Exp. \cite{pdg} &   $133\pm14$ &\multicolumn{3}{c}{$\pi\pi/K\bar K=0.39\pm0.14$}&&&&&&
\end{tabular}
\label{tab6}
\end{table}
\newpage
\centerline{Fig. 1}
\centerline{\epsfbox{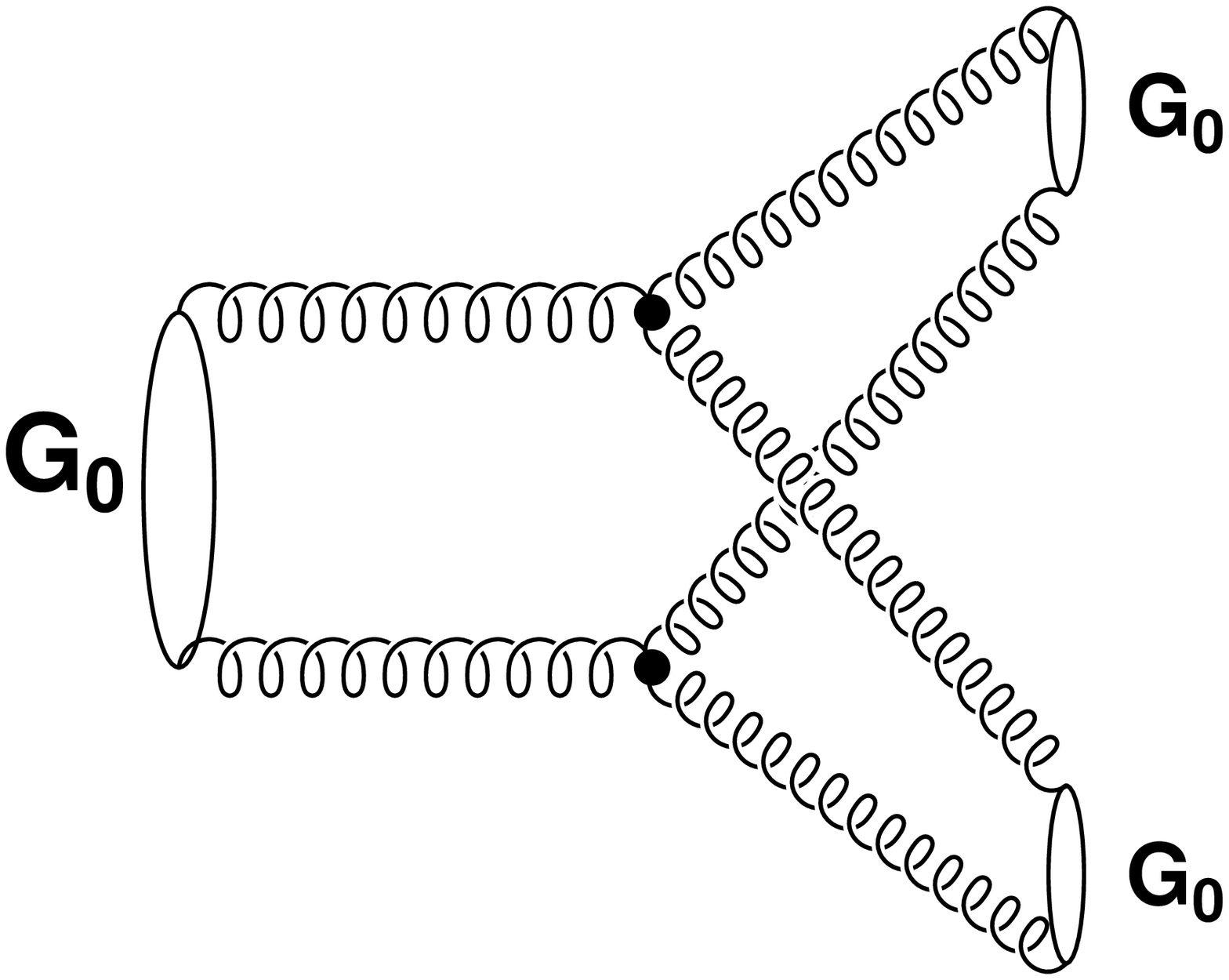}}
\newpage
\centerline{Fig. 2}
\centerline{\epsfbox{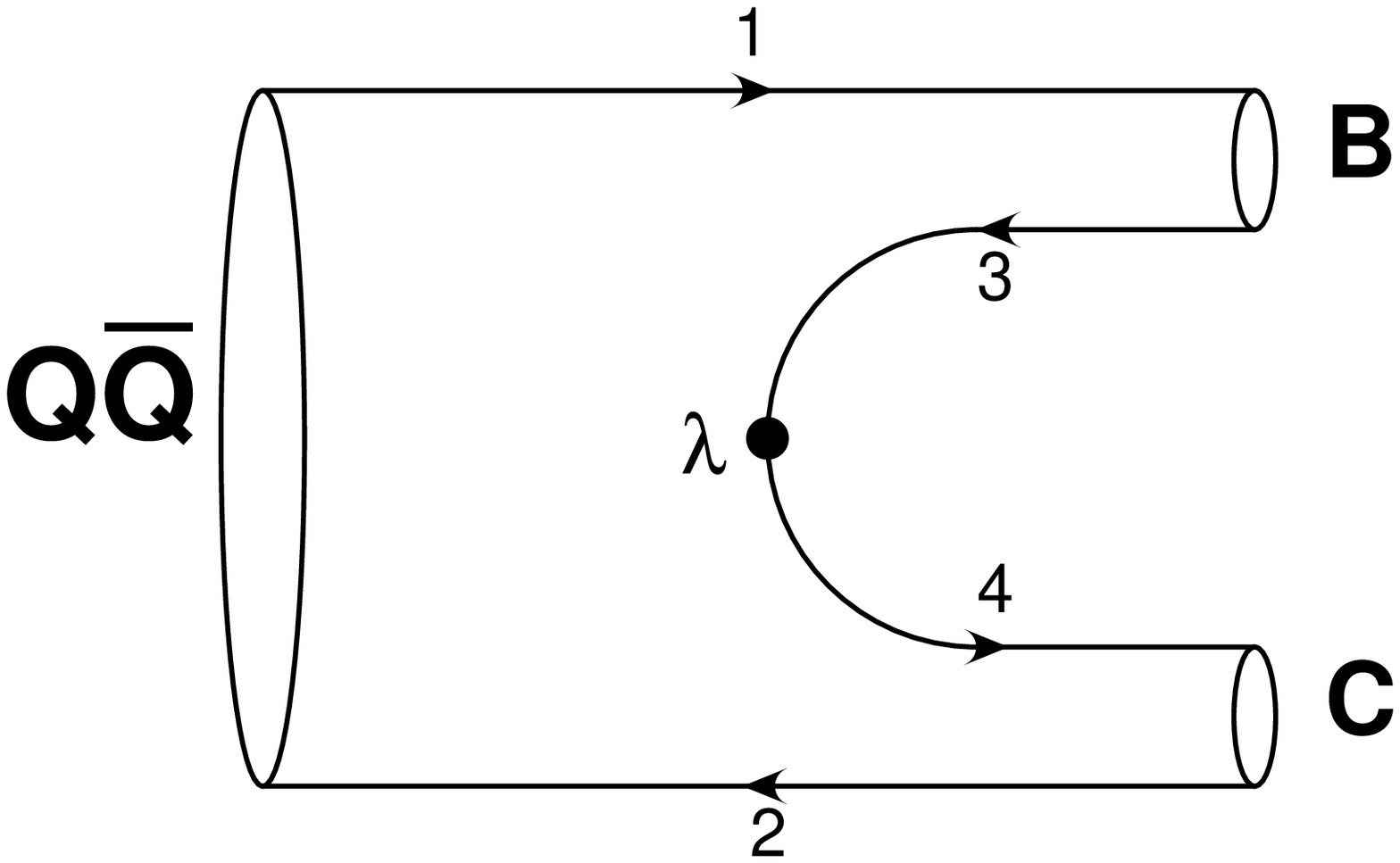}}
\newpage
\centerline{Fig. 3}
\centerline{\epsfbox{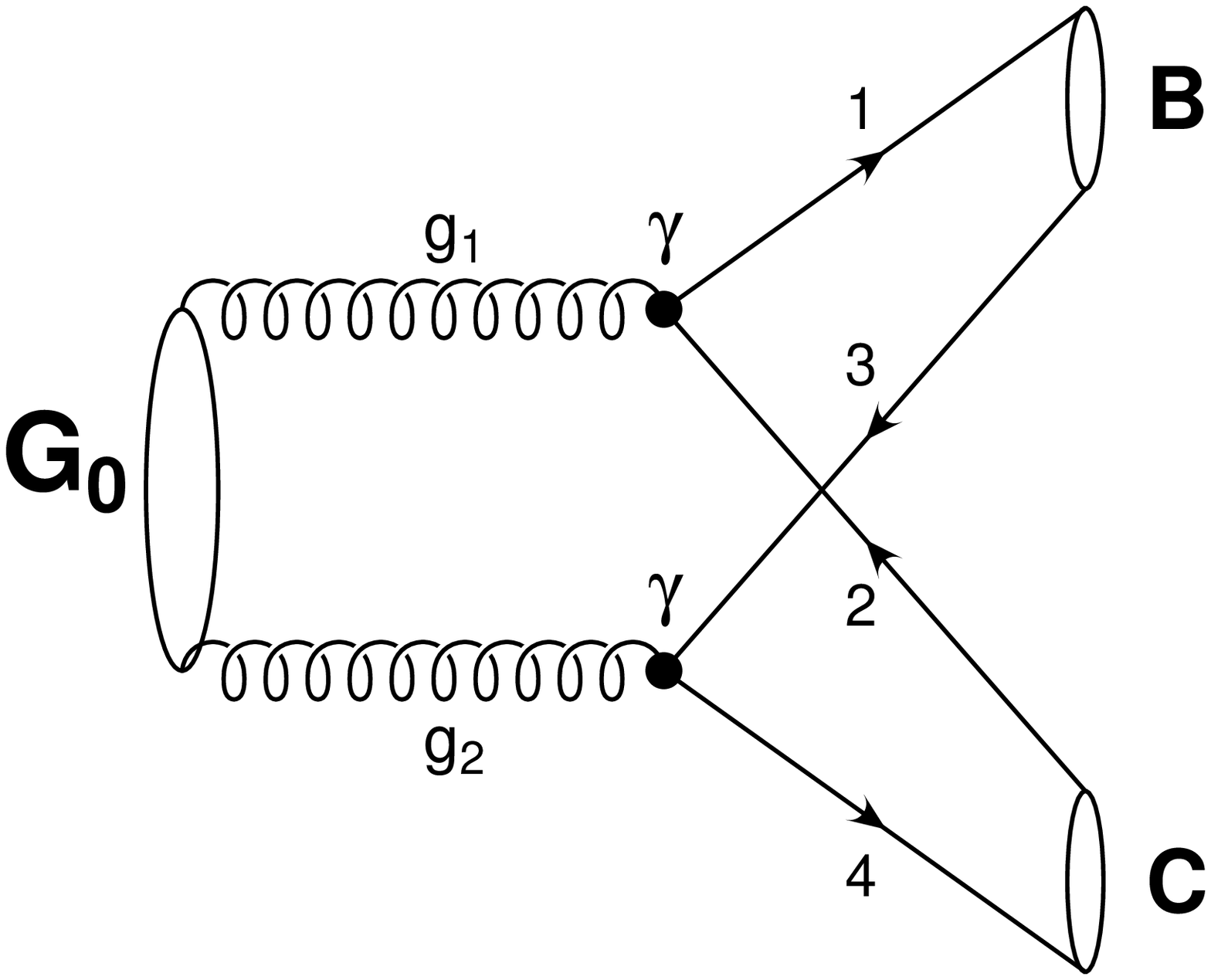}}
\newpage
\centerline{Fig. 4}
\centerline{\epsfbox{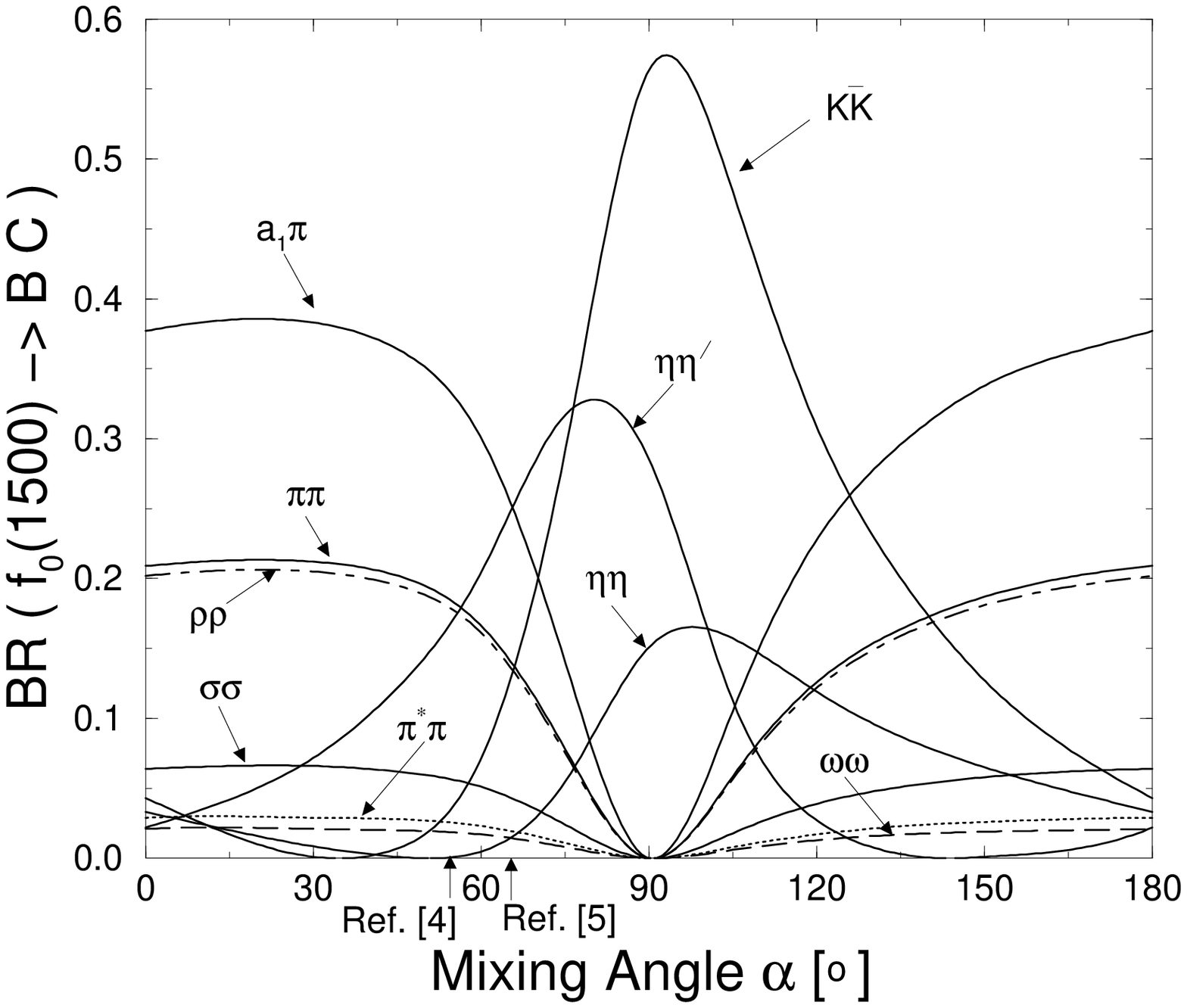}}
\newpage
\centerline{Fig. 5}
\centerline{\epsfbox{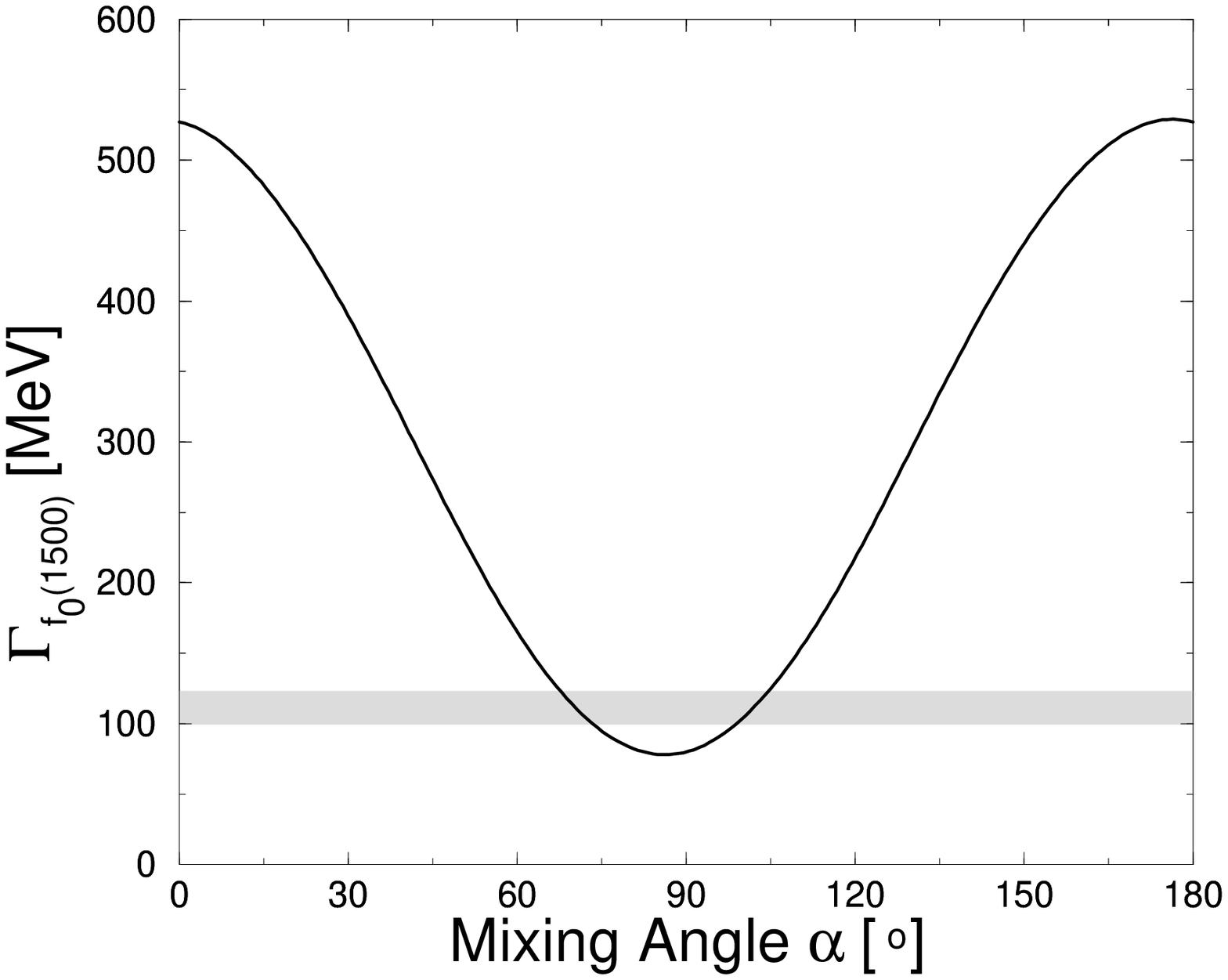}}
\newpage
\centerline{Fig. 6}
\centerline{\epsfbox{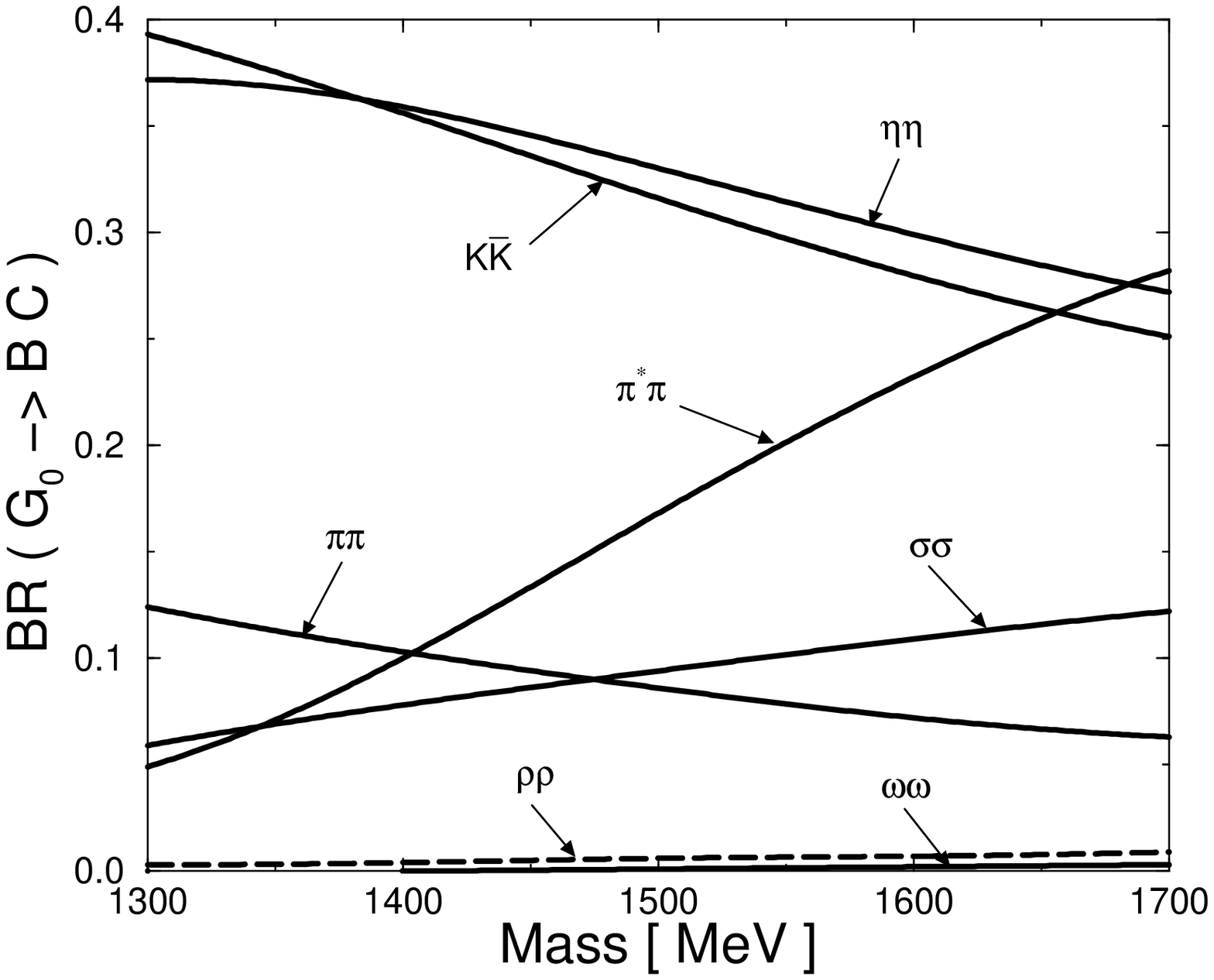}}
\newpage

\begin{thebibliography}{99}
\bibitem{clos} F.E. Close, Rep. Prog. Phys. 51, 833 (1988).
\bibitem{bal} G. Bali et al., Phys. Lett. B309, 378 (1993);
H. Chen, J. Sexton, A. Vaccarino and D. Weingarten, Nucl. Phys. B
(Proc. Suppl.) 34, 357 (1994).
\bibitem{tep} M. Teper,  hep-ph/9711299.
\bibitem{ams} C. Amsler and F.E. Close, Phys. Lett. B353, 385 (1996);
Phys. Rev. D53, 295 (1996); \\
F.E. Close, Nucl. Phys. B (Proc. Suppl.) 56A, 248 (1997).
\bibitem{weina} D. Weingarten, Nucl. Phys. B (Proc. Suppl.) 53, 232 (1997).
\bibitem{anis} A.V. Anisovich, V.V. Anisovich and A.V. Sarantsev,
Phys. Lett. B395, 123 (1997).
\bibitem{naris} S. Narison, Nucl. Phys. B509, 312 (1998).
\bibitem{amsler} C. Amsler, Rev. Mod. Phys. 70, 1293 (1998). \\
See also:  D. V. Bugg, A. V. Sarantsev, hep-ex/97120007.
\bibitem{bugg1710} D. V. Bugg et al., Phys. Lett. B353, 378 (1995);
W. Dunwoodie, SLAC-PUB-7163, Proc. of Hadron 97 (Upton N. Y. August 1997),
 p.753;
S. J. Lindenbaum  and R. S. Longacre, Phys. Lett. B274, 493 (1992).
\bibitem{pdg}  C. Caso et al.,
 The European Physical Journal C3, 1 (1998); R. M. Barnett et al., 
Phys. Rev. D54, 1 (1996).
\bibitem{kok} R. Kokoski and N. Isgur, Phys. Rev. D35, 907 (1987).
\bibitem{god}  Godfrey and N. Isgur, Phys. Rev. D32, 189 (1985).
\bibitem{mitja} M. Strohmeier-Pre\v{s}i\v{c}ek, T. Gutsche, R. Vinh Mau
and  Amand Faessler, Phys. Lett. B438, 21 (1998). 
\bibitem{weinb} W.Lee and D. Weingarten,  hep-lat/9805029.
\bibitem{thoma} U. Thoma, Nucl. Phys. B (Proc. Suppl.) 56A, 216 (1997).
\bibitem{meson} E. S. Ackley and T. Barnes, Phys. Rev. D54, 6811 (1996); \\
P. Geiger and E. S. Swanson, Phys. Rev. D50, 6855 (1994).
\bibitem{le} A. Le Yaouanc et al., Hadron transitions in the quark model, 
Gordon and Breach, Amsterdam (1988).
\bibitem{mary} M. Maruyama, S. Furui and Amand Faessler, Nucl. Phys. A472, 643 (1987)
\bibitem{nag} M. Nagels et al., Phys. Rev. D12, 744 (1975).
\bibitem{ulrike} U. Thoma , Proc. of Hadron 97 (Upton N. Y. August 1997),
p. 322.
\bibitem{pat} N. Isgur and J. Paton, Phys. Rev. D31, 2910 (1985).
\bibitem{dosch} H. G. Dosch and D. Gromes, Phys. Rev. D33, 1278 (1986).
\bibitem{thomas} T. Gutsche et al., to be published.
\bibitem{ps} C. Amsler et al., Phys. Lett. B294, 451 (1992).
\bibitem{jaffe} R. L. Jaffe and K. Johnson, Phys. Lett. 60B, 201 (1976);
Phys. Rev. Lett. 34, 1645 (1976).
\bibitem{iddir} A. Le Yaouanc et al., Z. Phys. C28, 309 (1985);
F. Iddir et al., Phys. Lett. B205, 564 (1988).
\bibitem{mosh} M. Moshinsky, The harmonic oscillator in modern 
physics: from atoms to quarks, Gordon and Breach, New York (1969);
T. A. Brody and M. Moshinsky, Tables of transformation brackets, Gordon and 
Breach, New York (1967). 
\bibitem{bugga} D. V. Bugg, A. V. Sarantsev, B. S. Zou, Nucl. Phys. B471, 59 (1996); 
 A. Abele et al., Nucl. Phys. A609, 562 (1996).
\bibitem{sexton} J. Sexton, A. Vaccarino and D. Weingarten,
 Phys. Rev. Lett. 75, 4563 (1995); Nucl. Phys. Proc. Suppl. 47, 128 (1996).
\bibitem{anisovich} V. V. Anisovich et al., Phys. Lett. B364, 195 (1995).
\bibitem{jin} H. Jin and X. Zhang, hep-ph 19805412.
\end{thebibliography}
\end{document}